\documentclass[amsmath,amssymb,aps,prd,showkeys,reprint, %
floatfix,groupedaddress,superscriptaddress,amsmath,amssymb]{revtex4-1}

\usepackage[breaklinks = true,colorlinks = true,%
linkcolor = blue,citecolor = blue]{hyperref}
\usepackage{graphicx}
\usepackage{xcolor}

\newcommand{\diag}{\mathrm{diag}}
\newcommand{\fm}{\;\text{fm}}
\newcommand{\MeV}{\;\text{MeV}}
\newcommand{\keV}{\;\text{keV}}

\newcommand{\bsigma}{\boldsymbol{\sigma}}
\newcommand{\bp}{\boldsymbol{p}}

\newcommand{\calC}{\mathcal{C}}
\newcommand{\calI}{\mathcal{I}}
\newcommand{\calG}{\mathcal{G}}
\newcommand{\calH}{\mathcal{H}}
\newcommand{\calM}{\mathcal{M}}
\newcommand{\calN}{\mathcal{N}}
\newcommand{\calGam}{\varGamma}
\newcommand{\mun}{\mu_{\rm n}}
\newcommand{\mup}{\mu_{\rm p}}
\newcommand{\mue}{\mu_{\rm e}}
\newcommand{\muu}{\mu_{\rm u}}
\newcommand{\mud}{\mu_{\rm d}}
\newcommand{\nB}{n_{\rm B}}
\newcommand{\np}{n_{\rm p}}
\newcommand{\nn}{n_{\rm n}}
\newcommand{\nne}{n_{\rm e}}
\newcommand{\nnu}{n_{\rm u}}
\newcommand{\nd}{n_{\rm d}}
\newcommand{\mmu}{m_{\rm u}}
\newcommand{\mmd}{m_{\rm d}}
\newcommand{\mmp}{m_{\rm p}}
\newcommand{\mmn}{m_{\rm n}}

\begin{document}


\title{Continuity from neutron matter to two-flavor quark matter\\ with
  $^1 S_0$ and $^3 P_2$ superfluidity}

\author{Yuki Fujimoto}
\email{fujimoto@nt.phys.s.u-tokyo.ac.jp}
\affiliation{Department of Physics, The University of Tokyo, 
  7-3-1 Hongo, Bunkyo-ku, Tokyo 113-0033, Japan}

\author{Kenji Fukushima}
\email{fuku@nt.phys.s.u-tokyo.ac.jp}
\affiliation{Department of Physics, The University of Tokyo, 
  7-3-1 Hongo, Bunkyo-ku, Tokyo 113-0033, Japan}
\affiliation{Institute for Physics of Intelligence (IPI), 
  The University of Tokyo, 7-3-1 Hongo, Bunkyo-ku, Tokyo 113-0033, Japan}

\author{Wolfram Weise}
\email{weise@tum.de}
\affiliation{Physics Department, Technical University of Munich, 85748
Garching, Germany}
\affiliation{Department of Physics, The University of Tokyo, 
  7-3-1 Hongo, Bunkyo-ku, Tokyo 113-0033, Japan}

\begin{abstract}
This study is performed with the aim of gaining insights into the
possible applicability of the quark-hadron continuity concept, not
only in the idealized case of three-flavor symmetric quark matter, but
also for the transition from neutron matter to two-flavor quark
matter.  A key issue is the continuity between neutron superfluidity
and a corresponding superfluid quark phase produced by $d$-quark
pairing.  Symmetry arguments are developed and relevant dynamical
mechanisms are analyzed.  It is pointed out that the $^3P_2$
superfluidity in dense neutron matter has a direct analogue in the
$^3P_2$ pairing of $d$-quarks in two-flavor quark matter.  This
observation supports the idea that the quark-hadron continuity
hypothesis may be valid for such systems.  Possible implications for
neutron stars are briefly discussed.  
\end{abstract}
\maketitle

\section{Introduction}

Two decades ago a conceptual framework for a continuous connection
between hadronic and quark phases of dense matter described by quantum
chromodynamics (QCD) was suggested in
Ref.\,\cite{Schafer:1998ef}, based on the exact matching of symmetry
breaking patterns and low-lying excitations in both domains.  In a
similar context, for three-flavor matter, correspondences between
condensates of pairs of hadrons and quarks have been discussed in
Ref.\,\cite{Alford:1999pa}.  These are the foundations for what is
called the ``quark-hadron continuity'' of matter at high baryon
density.  A Ginzburg-Landau analysis shows that matter at sufficiently
low temperature goes through a smooth crossover from the hadronic to
the quark phase as one increases the baryon
density~\cite{Hatsuda:2006ps}.  Such a continuous crossover is also
realized in a three-flavor Nambu--Jona-Lasinio
model~\cite{Abuki:2010jq}.  These features are further borne out by
the spectral continuity of Nambu-Goldstone (NG)
modes~\cite{Yamamoto:2007ah} and vector mesons~\cite{Hatsuda:2008is}.
Recently the continuity of topological defects such as superfluid
vortices that appear both in the hadronic phase and the color-flavor
locked (CFL) phases have been under
discussion~\cite{Alford:2018mqj,Chatterjee:2018nxe,Cherman:2018jir}.
A state-of-the-art result based on emergent higher-form symmetry gives
a plausible explanation for the quark-hadron vortex continuity to hold
even beyond the Ginzburg-Landau
regime~\cite{Hirono:2018fjr,*Hirono:2019oup}.
Some supplemental arguments for the continuity can be found also in
the large-$N_C$ limit (with $N_C$ being the color number) where the
color-superconducting gap is suppressed: quarkyonic
matter~\cite{McLerran:2007qj} refers to such continuity or duality
between nuclear and quark matter.  Implications of quarkyonic matter
to neutron star physics have been discussed in
Ref.\,\cite{Fukushima:2015bda}.  For phenomenology in favor of
quarkyonic matter, see recent
works~\cite{McLerran:2018hbz,Jeong:2019lhv}.

Inspired by these theoretical developments, the continuity scenario is
now also being considered in the context of neutron stars.  Particular
examples are the phenomenological constructions of the dense matter
equation of state (EoS), with quark-hadron continuity taken into
account~\cite{Masuda:2012kf,Alvarez-Castillo:2013spa,Baym:2017whm,Baym:2019iky}.
Conversely, recent attempts to extract the neutron star EoS directly
from astrophysical observations, using different methods such as
machine learning and Bayesian
inference~\cite{Fujimoto:2017cdo,*Fujimoto:2019hxv,Ozel:2010fw,*Ozel:2015ykl,Steiner:2010fz,*Steiner:2012xt,Abbott:2018exr},
may provide a basis for judging the continuity hypothesis.

The above-mentioned continuity concept is so far primarily based on
idealized SU(3) flavor symmetric settings. In reality, the strange
($s$) quark in QCD is much heavier than the up ($u$) and the down
($d$) quarks, with a mass ratio $m_s/m_{u,d} \sim 30$.  It is
therefore more natural to consider isospin-symmetric two-flavor
systems rather than starting from three-flavor symmetry.

A prototype example of dense baryonic matter is realized in the
interior of neutron stars.  Their composition is dominated by neutrons,
accompanied by a few percent of protons in $\beta$-equilibrium.  In
the present work we focus on superfluidity in neutron stars (see,
e.g., Refs.\,\cite{Gezerlis:2014efa,Sedrakian:2018ydt} for a
review). Under the aspect of quark-hadron continuity, the following
issue arises: as one proceeds to high baryon densities, does neutron
superfluidity have a corresponding analogue at the quark level? The
neutrons undergo BCS pairing in a $^1S_0$ state at low baryon
densities, i.e., $\nB < 0.5 \,n_0$ (with $n_0 \simeq 0.16 \fm^{-3}$,
the saturation density of normal nuclear matter).  This type of
superfluid is believed to exist in the inner crust of neutron stars.
With increasing baryon density, neutron pairing in the $^3P_2$ state
starts to develop and becomes the dominant pairing mechanism for
$\nB > n_0$, inward-bound towards the neutron star core region.  This
realization of $^3P_2$ superfluidity is based on the observed pattern
of nucleon-nucleon ($NN$) scattering phase
shifts~\cite{Hoffberg:1970vqj,Tamagaki:1970ptp,*Takatsuka:1992ga}.
The phase shift of the $^1S_0$ partial wave changes sign from positive
to negative with increasing energy of the two nucleons, indicating
that the pairing interaction turns from attractive to repulsive with
increasing Fermi energy.  Consequently, pairing in the $^1S_0$ channel
is disfavored at high densities and taken over by pairing in the
$^3P_2$ channel.  This property is attributed to the significant
attraction selectively generated by the spin-orbit interaction in the
triplet $P$-wave with total angular momentum $J = 2$.  All other
isospin $I=1$ $S$- and $P$-wave  $NN$ phase shifts are smaller or
repulsive in matter dominated by neutrons.  Various aspects and
properties of $^3P_2$ superfluidity inside neutron stars, from its
role in neutron star cooling to pulsar glitches, are subject to
continuing explorations (see, e.g.,
Refs.\,\cite{Bedaque:2003wj,Masuda:2015jka} and
\cite{Watanabe:2017nzj}).  A recent advanced analysis of pairing in
neutron matter based on chiral effective theory (EFT) interactions
including three-body forces can be found in
Ref.\,\cite{Drischler:2016cpy}.

\begin{figure}
  \includegraphics[width=\columnwidth]{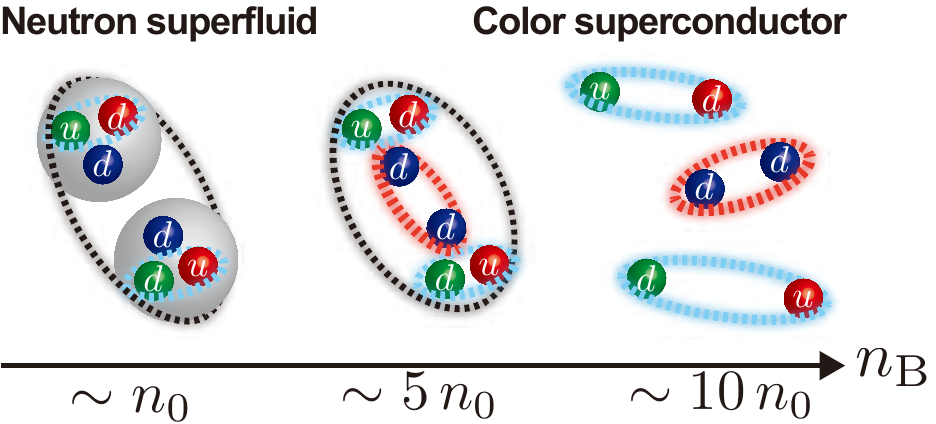}
  \caption{Schematic picture of quark-hadron continuity between
    neutron superfluid and color superconductor.  Cooper pairing of
    neutrons (indicated by dashed line) continuously connects to
    pairing of quarks in diquark condensates.}
  \label{fig:qhc}
\end{figure}

Our aim in this work is to investigate the continuity between superfluid
neutron matter and two-flavor quark matter with $^1 S_0$ and $^3 P_2$
superfluidity.  Related two-flavor NJL model studies have been
reported in Refs.\,\cite{Kitazawa:2003qmg,Zhang:2008ima}.  Here our
point is to collect and discuss the arguments which do indeed suggest
that the continuity concept applies to superfluid pairing when passing
from neutron matter to $u$-$d$-quark matter with a surplus of
$d$-quarks, as schematically illustrated in Fig.\,\ref{fig:qhc}.

This paper is organized as follows.  In Sec.~\ref{sec:neutron} we
describe some general physical properties of dense neutron star matter
and motivate the continuity between hadronic matter and quark matter
from a dynamical point of view.  Section~\ref{sec:symmetry} recalls
the conventional quark-hadron continuity scenario based on symmetry
breaking pattern considerations.  In Sec.~\ref{sec:operator}, we show
how the order parameter of $^3P_2$ neutron superfluidity can be
rearranged into two-flavor superconducting (2SC) $\langle ud \rangle$
and superfluid $\langle dd \rangle$ diquark condensates.
Section~\ref{sec:pwave} clarifies the microscopic mechanism that
induces the $\langle dd \rangle$ condensate in the $^3P_2$ state.  In
Sec.~\ref{sec:emt}, we demonstrate that the $^3P_2$
$\langle dd \rangle$ diquark condensate can be related to a
macroscopic observable, namely the pressure component of the
energy-momentum tensor.  This in turn is an important ingredient in
neutron star theories.  For an isolated nucleon it is also a key
subject of deeply-virtual Compton scattering measurements at
JLab~\cite{Burkert:2018bqq}.  In Sec.~\ref{sec:2scx}, discussions are
followed by a suggestive observation for the necessity of ``2SC+X'' to
fit the cooling pattern, where X may well be identified with the
$d$-quark pairing.  Finally, Sec.~\ref{sec:summary} summarizes our
findings.

\begin{figure}
  \includegraphics[width=\columnwidth]{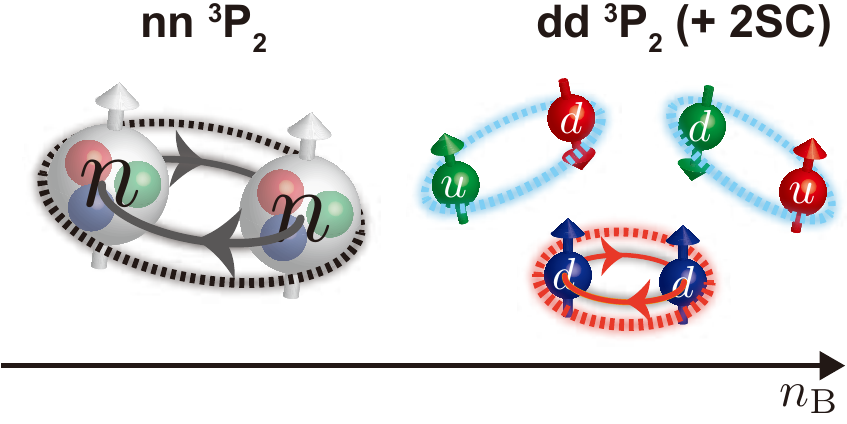}
  \caption{Schematic picture of quark-hadron continuity between the 
    $^3P_2$ neutron superfluid and the 2SC + $\langle dd \rangle$
    color superconductor.}
  \label{fig:qhc2}
\end{figure}

\section{Abundance of neutrons and down quarks in neutron star matter}
\label{sec:neutron}

In the extreme environment realized inside neutron stars, the
conditions of $\beta$-equilibrium and electric charge neutrality must
be satisfied.  A crude but qualitatively acceptable picture is that of
a degenerate Fermi gas of protons/neutrons and $u, d$
quarks.  Interaction effects will be taken into account later, but let
us first consider free particles and briefly overview the qualitative
character of the matter under consideration.
Here, we assume matter at densities around the onset of $u, d$ quarks where
  the onset of strangeness degrees of freedom may not occur yet.
This assumption is in accordance with the current two-solar-mass
pulsar constraints~\cite{Fonseca:2016tux, *Antoniadis:2013pzd,
  *Cromartie:2019kug}.

The $\beta$-equilibrium imposes a condition on the chemical potentials
of participating particles:
\begin{equation}
  \mun = \mup + \mue\,,\qquad
  \mud = \muu + \mue\,,
\label{eq:mu}
\end{equation}
for the hadronic and the quark phases, respectively. Here $\mue$ is
the chemical potential of the (negatively charged)
electrons.  Neutrinos decouple and do not contribute to the chemical
potential balance.  For a given baryon number density, $\nB$, in the
hadronic phase, we have two more conditions for the baryon number
density and the electric charge neutrality, namely,
\begin{equation}
  \np + \nn = \nB\,,\qquad  \np = \nne\,.
\label{eq:den} 
\end{equation}
For non-interacting particles the density is related to the chemical
potential through
\begin{equation}
  n_i = \frac{(\mu_i^2-m_i^2)^{3/2}}{3\pi^2}\,,
  \label{eq:density}
\end{equation}
where $i$ stands for p, n, e in the hadronic phase and for u, d, e in
the quark phase.  The equations
(\ref{eq:mu},\ref{eq:den},\ref{eq:density}) can then be solved for the
three variables, $\mup$, $\mun$, $\mue$, as functions of baryon
density $\nB$.

\begin{figure}
  \includegraphics[width=\columnwidth]{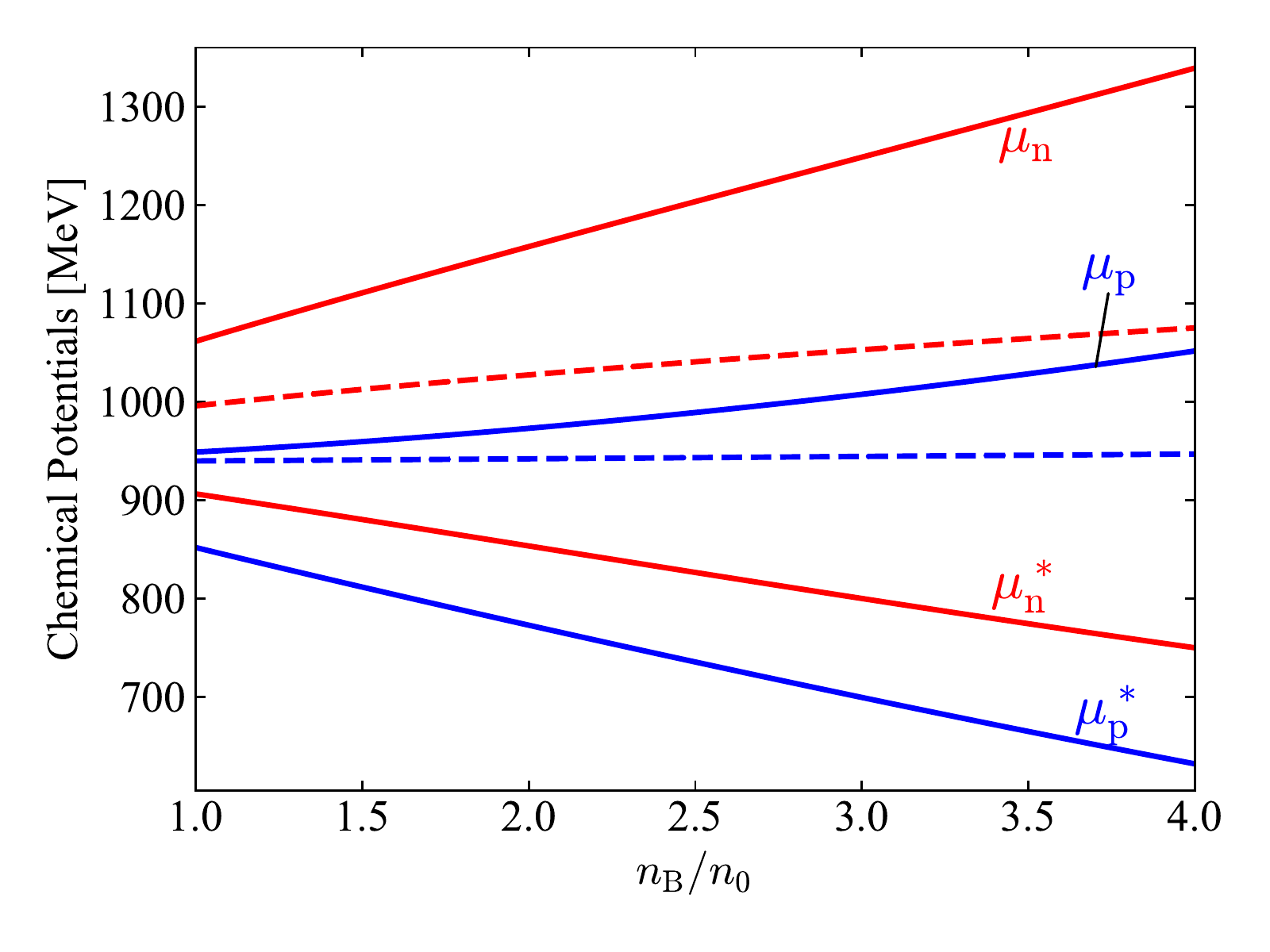}
  \caption{Nucleon chemical potentials, $\mup$, $\mun$, $\mup^\ast$,
    and $\mun^\ast$ as functions of the baryon number density $\nB$
    normalized by the normal nuclear density $n_0$.  The solid and
    dashed lines represents results with and without the interaction
    effects, respectively.}
  \label{fig:chem_n}
\end{figure}

In a relativistic mean-field picture of strongly interacting matter
the interaction effects are incorporated in terms of scalar and vector
condensates.  The scalar mean field changes the nucleon mass from its
vacuum value to a (reduced) in-medium effective mass.  The vector mean
field shifts the chemical potentials.  Here we are not interested in
fine-tuning parameters but rather in qualitative features of the Fermi
surface mismatch between different particle species.  With inclusion
of interactions, Eq.\,\eqref{eq:density} is modified with $\mu_i$
replaced by the shifted chemical potentials and $m_{\rm p/n}$ by the
in-medium masses:
\begin{align}
  &\mup^\ast = \mup - (G_v + G_\tau)\np - (G_v - G_\tau)\nn\,, \\
  &\mun^\ast = \mun - (G_v + G_\tau)\nn - (G_v - G_\tau)\np\,,\\
  &m_{\rm p/n}^\ast = m_{\rm p/n}\langle\sigma\rangle/f_\pi\,,
\end{align}
where $G_v$ and $G_{\tau}$ denote the coupling strength parameters of
isoscalar and isovector vector fields.  For guidance we use typical
couplings as they emerge in a chiral meson-nucleon field theory
combined with functional renormalization group methods, applied to
dense nuclear and neutron matter~\cite{Drews:2016wpi}:
\begin{equation}
  G_v \sim 4\fm^2\,,\qquad G_\tau \sim 1\fm^2\,.
\end{equation}
The scalar mean field $\langle\sigma\rangle$ is normalized to the pion
decay constant $f_\pi\simeq 92$ MeV in vacuum and decreases with
increasing baryon density.  Its detailed density dependence is
non-linear, but for the present discussion it is sufficient to realize
that $\langle\sigma\rangle$ drops to about half of its vacuum value at
$\nB \sim 5\,n_0$ (see Fig.\,25 of Ref.\,\cite{Drews:2016wpi}).  So we
parametrize the density dependence of the scalar condensate as
\begin{equation}
  \langle\sigma\rangle_\mu \simeq
  \langle\sigma\rangle_0 \Bigl(1-0.1\frac{\nB}{n_0}\Bigr)\,.
\end{equation}
Next we determine $\mup$, $\mun$, $\mue$ as functions of $\nB$.  The
energy dispersion relations are characterized by the in-medium quantities
$\mu_i^\ast$ and $m_i^\ast$.  The shifted chemical potentials are
shown in Fig.\,\ref{fig:chem_n}.  Solid lines represent results with
inclusion of the interaction effects using the parameters mentioned.
The dashed lines are the results with interactions turned off, i.e.,
using vacuum masses and no shifts on the chemical potentials.  In the
neutron star environment, $\mun^\ast$ is naturally larger than
$\mup^\ast$: neutrons dominate the state of matter.  Interestingly,
the Fermi surface mismatch between neutrons and protons is quite
stable with respect to interaction effects, while $\mue$ is
significantly modified.

\begin{figure}
  \includegraphics[width=\columnwidth]{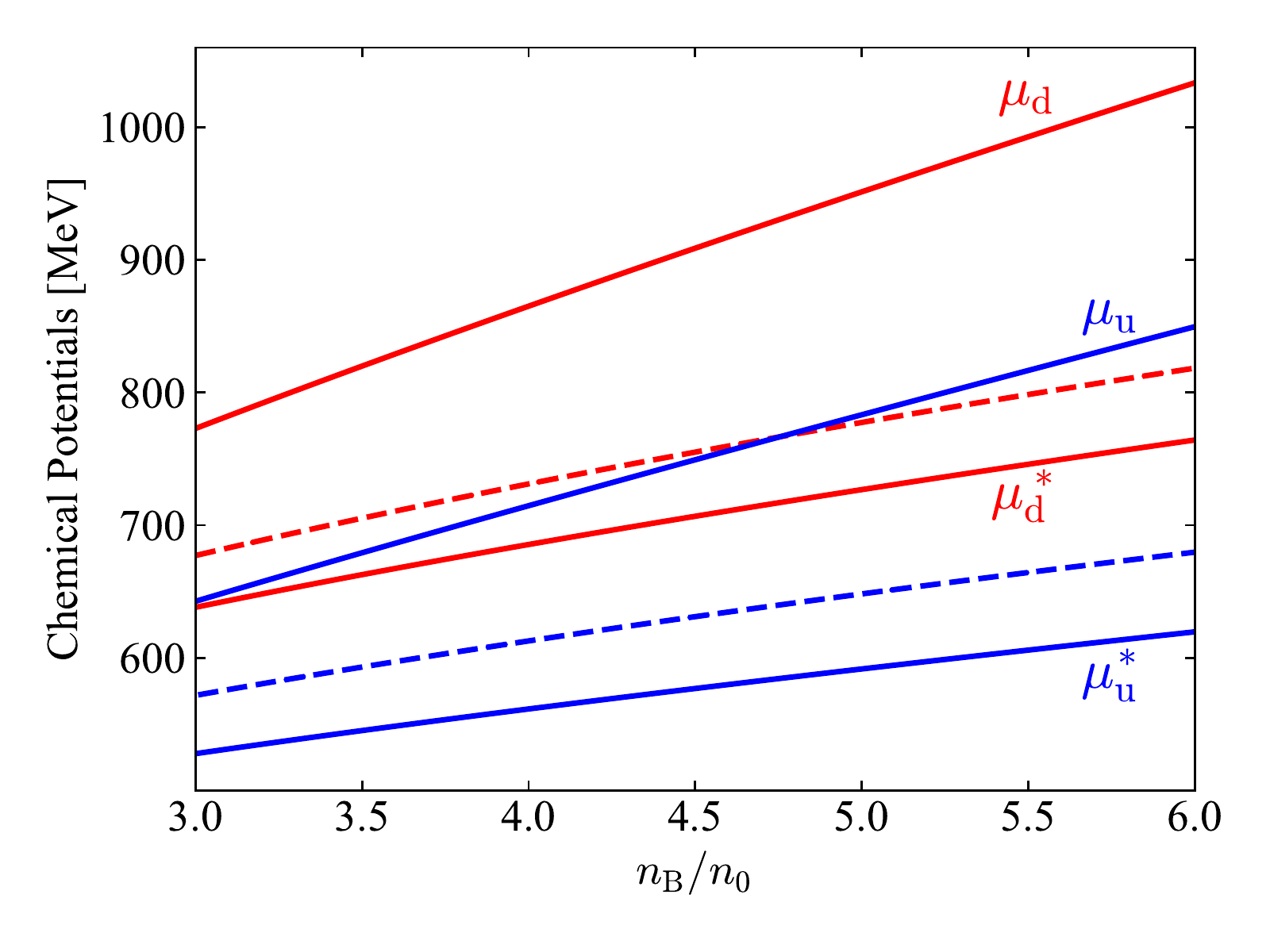}
  \caption{Quark chemical potentials, $\muu$, $\mud$, $\muu^\ast$, and
    $\mud^\ast$ as functions of the baryon number density $\nB$
    normalized by the normal nuclear density $n_0$.  The solid and
    dashed lines represents results with and without the interaction
    effects, respectively.}
  \label{fig:chem_q}
\end{figure}

For quark matter, the corresponding quark chemical potentials are
determined by an analogous set of three conditions.  Apart from
binding energy effects which we neglect here for simplicity, we use
constituent quark masses,
\begin{equation}
  \mmu = 312.3\MeV\,,\qquad \mmd = 313.6\MeV\,,
\end{equation}
fixed to reproduce physical proton and neutron masses,
$\mmp=2\mmu+\mmd$ and $\mmn=2\mmd+\mmu$.   We note that the vector
couplings $g_v$ and $g_\tau$ in the quark sector should be smaller
than $G_v$ and $G_\tau$ by $1/9$ because of the difference by a factor
$N_{C} =3$ between baryon and quark number.  It is an interesting
observation that our input, $g_v\sim G_v/9 = 0.44\fm^2$, is
suggestively close to a recent estimate \,\cite{Song:2019qoh}:
$g_v\sim \pi\alpha_{\rm s}/(3p_{\rm F}^2) \sim 0.5\fm^2$ (an
additional factor of two appears here  because of a different
convention in Ref.\,\cite{Song:2019qoh}).  For the density-dependent
constituent quark masses we assume the same scaling with
$\langle\sigma\rangle$ as for the nucleon mass.  In-medium chemical
potentials and quark masses are then incorporated as
\begin{align}
  &\muu^\ast = \muu - (g_v+g_\tau) \nnu - (g_v-g_\tau)\nd\,,\\
  &\mud^\ast = \mud - (g_v+g_\tau) \nd - (g_v-g_\tau)\nnu\,,\\
  &m_{\rm u/d}^\ast = m_{\rm u/d} \langle\sigma\rangle/f_\pi\,.
\end{align}
Figure~\ref{fig:chem_q} shows the shifted quark chemical potentials as
functions of $\nB$.  In this case again, $\mud^\ast$ is naturally
larger than $\muu^\ast$ for neutron-rich matter in $\beta$-equilibrium
and under the electric neutrality condition.  At high baryon densities
this Fermi surface mismatch between $d$ and $u$ quarks shows a
correspondence to the mismatch between neutrons and protons in neutron
star matter.  It suggests the possibility of pairing in the $I=1$ $dd$
channel analogous to the superfluid neutron pairing mentioned
previously.

\section{Symmetry arguments for quark-hadron continuity}
\label{sec:symmetry}

Here we give a brief overview of quark-hadron continuity from the
symmetry point of view.  If the pattern of spontaneous symmetry
breaking features a discontinuity between two states or compositions
of matter, there must be at least one phase transition separating
these two states.  This implies that, if two such states are smoothly
connected without a phase transition, the symmetry breaking pattern
must be identical on both sides. We describe in the following how this
symmetry argument works for quark-hadron continuity, first in the
three-flavor case and next in the two-flavor case.  While the former
is well established through the pioneering work of
Ref.\,\cite{Schafer:1998ef}, the latter is a novel scenario that we
are proposing in the present work.

\subsection{Three-flavor case}
\label{sec:three}

The ground state of three-flavor symmetric quark matter at high
density supposedly accommodates diquark condensates featuring a CFL
phase.  It has been demonstrated that the CFL phase is characterized
by the same symmetry breaking pattern as the hadronic phase with a
superfluid~\cite{Schafer:1998ef}.  Here, diquarks in the
color-antitriplet, the flavor-triplet, and the scalar channel, which
are often called the ``good'' diquarks in the context of exotic
hadrons (see, e.g., Ref.\,\cite{Wilczek:2004im,*Jaffe:2004ph}), play
an essential role for the symmetry argument.  We thus introduce the
corresponding good diquark operator as
\begin{equation}
  \hat{\Phi}^{\alpha A} \equiv \calN\,\epsilon^{\alpha\beta\gamma}
  \epsilon^{ABC} \,\hat{q}_{\beta B}^\top \,{\cal C}\gamma_5\, \hat{q}_{\gamma C}\,,
  \label{eq:good}
\end{equation}
where $\cal N$ is a normalization~\footnote{Strictly speaking,
  Eq.~\eqref{eq:good} is only valid for the good diquark.  Two
  quarks in $\boldsymbol{3}$ representation of color and flavor group
  reduce to $\boldsymbol{3} \times \boldsymbol{3} =
  \bar{\boldsymbol{3}} + \boldsymbol{6}$. The indices on the right-hand
  side of Eq.~\eqref{eq:good} are those of $\boldsymbol{3}$, while on
  the left-hand side they refer to $\bar{\boldsymbol{3}}$ after the
  reduction.  See also Eqs.~\eqref{eq:CFLcond} and \eqref{eq:2SCcond}.}.
In the present study numerical
values of superconducting gaps are not essential, so we often omit the
normalization factor for simplicity.  The charge conjugation matrix,
${\cal C}\equiv i\gamma^0\gamma^2$, is inserted to form a Lorentz
scalar.  In the expression above the spin or Dirac indices are all
contracted implicitly.  Greek indices ($\alpha,\beta,\gamma$) and
capital indices ($A,B,C$) represent color and flavor, respectively.

In terms of left-handed and right-handed fermions, the diquark
operator can be decomposed into $\hat{\Phi}_{\rm L}^{\alpha A}$ and
$\hat{\Phi}_{\rm R}^{\alpha A}$, respectively.  Because diquark
condensation in the scalar channel is favored by the axial anomaly,
the left- and right-handed condensates,
$\Phi_{\rm L}^{\alpha A}\equiv\langle\hat{\Phi}_{\rm L}^{\alpha A}\rangle$
and $\Phi_{\rm R}^{\alpha A}\equiv\langle\hat{\Phi}_{\rm R}^{\alpha A}\rangle$,
in the CFL phase have the property:
\begin{equation}
  \Phi_{\rm L}^{\alpha A} = -\Phi_{\rm R}^{\alpha A} = \delta^{\alpha A} \Delta\,,
  \label{eq:CFLcond}
\end{equation}
where gauge fixing is assumed so that the color direction aligns with
flavor as $\delta^{\alpha A}$, and $\Delta$ is a gap parameter.

Clearly $\Phi_{\rm L}^{\alpha A}$ breaks both flavor
$\mathrm{SU(3)_L}$ and color $\mathrm{SU(3)_C}$, but a simultaneous
color-flavor rotation can leave $\Phi_{\rm L}^{\alpha A}$ unchanged.
In the same way $\Phi_{\rm R}^{\alpha A}$ breaks both flavor
$\mathrm{SU(3)_R}$ and color $\mathrm{SU(3)_C}$ down to their
vectorial combination.  This unbroken vectorial symmetry is commonly
denoted as $\mathrm{SU(3)_{C+L+R}}$.  Hence the symmetry breaking
pattern can be summarized as $\calG\to\calH$ with
\begin{equation}
  \begin{split}
  \calG &= [\mathrm{SU(3)_C}] \times \mathrm{SU(3)_L} \times
  \mathrm{SU(3)_R} \times \mathrm{U(1)_B}~, \\
  \calH &= \mathrm{SU(3)_{C+L+R}}~,
  \end{split}
  \label{eq:CFLbreaking}
\end{equation}
apart from redundant discrete symmetries.  Here $[\mathrm{SU(3)_C}]$
represents the global part of color symmetry (while local gauge
symmetry is never broken).  The spontaneous breaking of global color
symmetry makes all eight gluons massive due to the Anderson-Higgs
mechanism.  It is important to note that $\mathrm{U(1)_B}$
corresponding to baryon number conservation is spontaneously broken,
so that the CFL state can be regarded as a superfluid.  A more
detailed discussion on nontrivial realization of the $\mathrm{U(1)_B}$
breaking will be given when we consider the two-flavor case in what
follows.

The crucial point is now that chiral symmetry
breaking\,\eqref{eq:CFLbreaking} in the CFL phase is identical to the
familiar scenario in the hadronic phase.  The low-energy properties of
matter are governed by NG bosons, which implies that chiral EFT can be
systematically formulated for the CFL
state~\cite{Casalbuoni:1999wu,Son:1999cm,*Son:2000tu}.  Therefore the
theoretical descriptions of hadronic and CFL matter are analogous by
construction.  This is the basic message of
Ref.\,\cite{Schafer:1998ef} which pointed out the important
possibility that hadronic and CFL matter can be continuously and
indistinguishably connected.

Continuity is a strong hypothesis, requiring a one-to-one
correspondence between physical degrees of freedom in hadronic and
quark matter.  The CFL phase works with quarks, gluons and chiral NG
bosons.  The spectrum of their excitations can be translated into the
relevant composite degrees of freedom in the hadronic phase: nonet
baryons, octet vector mesons, and the octet of pseudoscalar NG bosons.
Further steps have recently been made investigating the issue of
vortex continuity but some controversies still remain.

From the discussions so far one may have thought that
$\mathrm{U(1)_B}$ is not necessarily broken in the hadronic phase.
Surely, on the one hand, the hadronic vacuum at zero density does not
break $\mathrm{U(1)_B}$.  On the other hand, it is known that nuclear
matter can have a superfluid component generated by the pairing
interaction of nucleons.  It is thus conceivable that superfluidity
also develops in idealized three-flavor symmetric baryonic matter.  We
shall return to related considerations in Sec.~\ref{sec:operator}
where a superfluid operator for baryons will be explicitly
identified.

\subsection{Two-flavor case}
\label{sec:two}

The color-flavor-locked configurations assign a special significance
to $N_F = N_C = 3$: quark-hadron continuity is usually not postulated
for the two-flavor case.  In this subsection we point out, however,
that such a continuity scenario is also possible for two-flavor
nuclear and quark matter.
In order for the two-flavor continuity scenario to make sense, the
requirements at the quark matter side are: (1) Strangeness is negligible,
(2) Quarks are deconfined and the chiral symmetry is still broken, (3)
Baryon superfluidity occurs.

  \subsubsection{2SC phase}

The ground state of two-flavor symmetric quark matter at high density
is considered to be the 2SC phase with the following condensates,
\begin{equation}
  \Phi_{\rm L}^{\alpha A} = -\Phi_{\rm R}^{\alpha A}
  = \delta^{\alpha 3} \delta^{A 3} \Delta\,.
  \label{eq:2SCcond}
\end{equation}
The color direction, $\delta^{\alpha 3}$, is a gauge choice consistent
with Eq.\,\eqref{eq:CFLcond}.  These condensates imply a symmetry
breaking pattern, $\calG\to\calH$, with
\begin{equation}
  \begin{split}
  \calG &= [\mathrm{SU(3)_C}] \times \mathrm{SU(2)_L} \times
  \mathrm{SU(2)_R} \times \mathrm{U(1)_B}\,, \\
  \calH &= [\mathrm{SU(2)_C}] \times \mathrm{SU(2)_L} \times
  \mathrm{SU(2)_R} \times \mathrm{U(1)_{C+B}}\,.
  \end{split}
\end{equation}
The 2SC condensates partially break the global color symmetry: five
out of eight gluons become massive.  Since the flavor structure of
Eq.\,\eqref{eq:2SCcond} is a singlet in the two-flavor sector, chiral
symmetry remains intact.  Moreover, a modified version of
$\mathrm{U(1)_B}$ survives unbroken.

To exemplify the unbroken $\mathrm{U(1)_{C+B}}$, consider the
color-flavor combinations of the pairing underlying
Eq.\,\eqref{eq:2SCcond}.  The 2SC phase has nonzero condensates,
\begin{equation}
  \langle (ru)(gd)\rangle\,,\quad
  \langle (rd)(gu)\rangle\,,
\end{equation}
where $(ru)$ denotes a red $u$ quark, etc.  Under the
$\mathrm{U(1)_B}$ transformation, $\hat{q}\to e^{i\theta/3}\hat{q}$,
these two pairs receive a phase $e^{2i\theta/3}$ which can be canceled
by a color rotation, $\hat{q}\to e^{-i(2/\sqrt{3})\theta T_8}\hat{q}$,
with $T_8=\frac{1}{2\sqrt{3}}\diag(1,1,-2)$.  In the same way we see
that the 2SC phase is not an   electromagnetic superconductor.  The
original $\mathrm{U(1)_{\rm em}}$ symmetry generated by
$Q_e=\diag(\frac{2}{3},-\frac{1}{3})e$ is broken, but modified
$\mathrm{U(1)_{\rm \widetilde{em}}}$ remains unbroken which is
generated by a mixture of $Q_e$ and $T_8$,
\begin{equation}
  \tilde{Q}_e = Q_e - \frac{e}{\sqrt{3}} T_8\,.
  \label{eq:modQe}
\end{equation}
It is therefore evident that the pure 2SC phase itself cannot be
smoothly connected to the hadronic phase: symmetry breaking patterns
are different.  Nevertheless, a coexisting phase is not excluded, in
which a chiral condensate $\langle\bar{q}q\rangle$ and diquark
condensates~\eqref{eq:2SCcond} are simultaneously non-zero.
Coexistence has been confirmed in the preceeding model calculations
in Refs.~\cite{Berges:1998rc,*Huang:2001yw}.  Hereafter we assume
$\langle\bar{q}q\rangle\neq 0$ in our following discussions.  In this
way the chiral symmetry breaking part is trivially matched to the
hadronic phase.  Below we see that this assumption can be relaxed by
an additional condensate.

In contrast to chiral symmetry broken by $\langle \bar{q}q\rangle$,
superfluidity is a nontrivial issue.  As previously mentioned, the
hadronic phase has a superfluid component generated by pairing
interactions between  nucleons.  The quark matter analogue should
therefore likewise break $\mathrm{U(1)_B}$ in order for the continuity
scenario to be consistently valid.

  \subsubsection{2SC$+\langle dd \rangle$ phase}

As discussed in Sec.~\ref{sec:neutron}, neutron matter with its
maximal isospin asymmetry has an abundance of $d$ quarks which are not
paired with $u$ quarks.  One can therefore anticipate the formation of
a $\langle dd\rangle$ diquark condensate at high baryon densities.
The microscopic structure of $\langle dd\rangle$ will be clarified
later; for the moment let us consider the simplest case, namely,
scalar $\langle dd\rangle$ in the color-sextet channel.  On first
sight such a condensate appears not to be favored because the
one-gluon exchange interaction in the color-sextet channel is
repulsive.  But it will turn out as we proceed that this repulsive
short-distance force is important for the microscopic structure of
$\langle dd\rangle$.

Now, if a non-zero $\langle dd\rangle$ in the color-sextet channel
coexists in the 2SC phase which may well be called the
2SC+$\langle dd\rangle$ phase, we can confirm that
$\mathrm{U(1)_B}$ symmetry or its modified variants do not survive.
The possible color-flavor combinations are,
\begin{equation}
  \langle (\alpha d)(\beta d)\rangle\,,
\end{equation}
where the color pairs are symmetric:
$(\alpha,\beta)=(r,r)$, $(g,g)$, $(b,b)$, $(r,g)$, $(g,b)$, $(b,r)$.
Under the transformation,
$\hat{q}\to e^{i\theta} e^{-i2\sqrt{3}\theta T_8}\hat{q}$, the pairs
$(\alpha,\beta)=(r,r)$, $(g,g)$, $(r,g)$ are invariant, but the remaining 
three combinations change nontrivially.  If we consider continuity from
neutron matter, $(\alpha,\beta)=(b,b)$ is favored since $ud$ diquarks
are chosen as Eq.\,\eqref{eq:2SCcond} in a gauge-fixed description of
the 2SC phase. Thus, the 2SC+$\langle dd\rangle$ phase breaks
$\mathrm{U(1)_B}$ and exhibits superfluidity.  Also, $\langle dd
\rangle$ induces the chiral symmetry breaking even without the chiral
condensate.  This $\langle dd \rangle$ fulfills the desired properties
for the quark-hadron continuity to be valid, which are lacking in the
pure 2SC phase.
The dynamical aspect of the chiral symmetry breaking in a certain
model deserves further consideration as a future work.
Here we note that the single-color and single-flavor
pairing such as $\langle(bd)(bd)\rangle$ has been studied in
the preceding work~\cite{Alford:2002rz}.

Finally, before closing our symmetry argument for quark-hadron
continuity, we note that modified electromagnetic
$\mathrm{U(1)_{\widetilde{\rm em}}}$ remains unbroken, so the
2SC+$\langle dd\rangle$ phase cannot be an electromagnetic
superconductor.  To confirm this, the quickest way is that $(bd)$
quarks, dominant constituents in $\langle dd\rangle$, are neutral with
respect to $\tilde{Q}_e$.  Therefore, $\langle dd\rangle$ does not
affect the $\mathrm{U(1)_{\widetilde{\rm em}}}$ symmetry.
The charge properties of the $(bu)$ and $(bd)$ quarks in the 2SC were
explicitly given in Ref.\,\cite{Alford:2014doa}.

In the CFL phase, $(bu)$ and $(bd)$ quarks are identified with protons
and neutrons, respectively~\cite{Alford:1999pa}, thus it is also
natural to expect the neutron condensate $\langle nn\rangle$ maps to
$\langle (bd)(bd)\rangle$ condensate in the 2SC phase. It is
also worth mentioning that $\langle (bu)(bu)\rangle$
breaks the $\mathrm{U(1)_{\widetilde{\rm em}}}$ symmetry, which is
consonant with the fact that the $\langle pp\rangle$ condensate
induces the proton superconductivity.

Even with $\langle dd \rangle$ condensation, there remain unpaired
quarks in the 2SC+$\langle dd \rangle$ phase.  These unpaired quarks
do not affect the continuity but may dominate low energy excitations,
which may eventually be suppressed by dynamical symmetry breaking.

\section{Rearrangement of the order-parameter operators}
\label{sec:operator}

The following exercise is to formally demonstrate quark-hadron
continuity using gauge-invariant order parameters.  An essential
observation for the intuitive understanding of quark-hadron continuity
lies in the fact that no physical or gauge-invariant quantity can
discriminate nuclear and quark matter.  This observation is traced
back to the absence of any order parameter for deconfinement of
dynamical quarks in the color fundamental representation.

Throughout this work we describe the low-lying baryons in terms of a
quark-diquark structure;  for our purpose matching of the right
quantum number is sufficient.
In this picture the colorless spin-$\frac{1}{2}$ baryon
operators, with flavor indices $A,B$ shown explicitly, are given by:
\begin{equation}
  \hat{{\cal B}}^{AB}_\sigma = \hat{\Phi}^{\alpha A} \hat{q}^B_{\alpha \sigma}\,,
\end{equation}
where $\sigma$ denotes the spin index.  We note again that the
normalization is dropped for notational brevity.
This baryon interpolating operator may well have the largest overlap with the
physical state, so such a combination of quark-diquark can be regarded
as a reasonable approximation for baryon wave-functions.
In any case, as long as we consider the quark-hadron
continuity, what really matters is the quantum number only.

\subsection{Three-flavor symmetric case}

Here we start with the order parameters in the CFL phase which are
then translated into the hadronic representation.  The gauge-invariant
order parameters are the mesonic and the baryonic condensates defined
as
\begin{align}
  \calM^{AB} &= \langle\hat{\calM}^{AB}\rangle
  = \langle \hat{\Phi}^{\dagger A \alpha}
  \hat{\Phi}^{\alpha  B} \rangle\,,\\
  \Upsilon^{ABC} &= \langle\hat{\Upsilon}_{\rm CFL}^{ABC}\rangle
  = \langle \epsilon^{\alpha\beta\gamma}
  \hat{\Phi}^{\alpha A} \hat{\Phi}^{\beta B} \hat{\Phi}^{\gamma C} \rangle\,,
\end{align}
respectively.  We are primarily interested in superfluidity aspects
and hence focus on $\hat{\Upsilon}^{ABC}$.  Decomposing
$\hat{\Phi}^{\alpha A}$ into quarks and combining the quark operators
with the remaining $\hat{\Phi}^{\beta B}$ and $\hat{\Phi}^{\gamma C}$
to form two-baryon operators, one arrives at
\begin{equation}
  \hat{\Upsilon}^{ABC} = 2\epsilon^{AMN}
  \hat{\cal B}^{BM}_\sigma ({\cal C}\gamma_5)_{\sigma\sigma'} \hat{\cal B}^{CN}_{\sigma'} \,.
\end{equation}
Let us now limit ourselves to the octet baryons:
\begin{equation}
  {\cal B}_{\boldsymbol{8}}^{AB} = \begin{pmatrix}
    \frac{1}{\sqrt{2}}\Sigma^0 + \frac{1}{\sqrt{6}}\Lambda &
    \Sigma^+ & p \\
    \Sigma^- & -\frac{1}{\sqrt{2}}\Sigma^0 + \frac{1}{\sqrt{6}}\Lambda & n \\
    \Xi^- & \Xi^0 & -\frac{2}{\sqrt{6}}\Lambda
  \end{pmatrix}_{AB}\,,
\end{equation}
where $({\cal C}\gamma_5)^\top=-{\cal C}\gamma_5$ is used with $\top$
denoting the transpose.  Thus, the flavor-singlet CFL order parameter,
$\Upsilon^{(0)} \equiv \epsilon^{ABC}\Upsilon^{ABC}$, is smoothly
connected to superfluid strange baryonic matter, explicitly
represented as
\begin{align}
  &\Upsilon^{(0)} = 2({\cal C}\gamma_5)_{\sigma\sigma'}\langle
    \hat{\cal B}^{AA}_{\boldsymbol{8} \sigma} \hat{\cal B}^{BB}_{\boldsymbol{8} \sigma'}
    - \hat{\cal B}^{AB}_{\boldsymbol{8} \sigma} \hat{\cal B}^{BA}_{\boldsymbol{8} \sigma'}  \rangle \notag\\[0.5em]
  & \propto ({\cal C}\gamma_5)_{\sigma\sigma'}
    \langle \tfrac{1}{2}\Lambda_\sigma \Lambda_{\sigma'} \!+\! \tfrac{1}{2}\Sigma^0_\sigma \Sigma^0_{\sigma'}
    \!+\! \Sigma^+_\sigma \Sigma^-_{\sigma'} \!+\! p_\sigma \Xi^-_{\sigma'} \!+\! n_\sigma \Xi^0_{\sigma'} \rangle\,.
\label{eq:non-rela-pairing}
\end{align}
At this point we consider the non-relativistic reduction of the
dibaryonic condensates.  Conventinally the ${}^3P_2$ neutron superfluidity has
been discussed in the non-relativistic regime, so it is
useful to see what the relativistic counterpart of the
non-relativistic condensates is.
The term $\langle\Lambda\Lambda\rangle$ may serve as a specific
example.  The generalization to other terms is straightforward.  Using
a solution of the Dirac equation with $\gamma^\mu$ in the Dirac
representation, the four-component spinor of the $\Lambda$ is
expressed as
\begin{equation}
  \Lambda = \begin{pmatrix}
    \varphi_\Lambda \\[0.6em]
    \displaystyle \frac{\bsigma\cdot\bp}{E_p+m_\Lambda}\varphi_\Lambda
  \end{pmatrix}\,,
  \label{eq:non-rela}
\end{equation}
with two-component spinors $\varphi_\Lambda$.  The lower components
are negligible in the limit $m_\Lambda \gg |\bp|$ and one finds
\begin{equation}
  ({\cal C}\gamma_5)_{\sigma\sigma'}\langle \Lambda_\sigma\Lambda_{\sigma'}\rangle
  = \langle\varphi_\Lambda^\top i\sigma^2 \varphi_\Lambda \rangle\,,
\label{eq:nr}
\end{equation}
in terms of the non-relativistic wave function $\varphi_\Lambda$ whose
two components correspond to the spin degrees of freedom.

\subsection{Two-flavor $^1 S_0$ superfluid matter}

The preceding subsection started by identifying the ground state as
quark matter in the CFL phase followed by the rearrangement of order
parameters in terms of baryonic operators.  For the two-flavor case we
follow an inverse sequence of steps: the starting point is now neutron
matter with a superfluid component and we investigate the possibility
of a continuous transition to superfluid quark matter with an excess
of $d$-quarks.

As long as the baryon density is below the onset of $P$-wave
superfluidity, the neutron superfluid occurs in the $^1 S_0$ channel.
In this case the superfluid order parameter in neutron matter is given
by $\langle\varphi_{\rm n}^\top i\sigma^2\varphi_{\rm n}\rangle$ in
the non-relativistic representation [see Eq.\,\eqref{eq:nr}], which
can be generalized into a relativistic expression as
\begin{equation}
  \langle \varphi_{\rm n}^\top i\sigma^2 \varphi_{\rm n} \rangle
  \;\;\to\;\;
  \langle\hat{\Upsilon}_S \rangle \equiv
  \langle \hat{n}_\sigma ({\cal C}\gamma_5)_{\sigma\sigma'} \hat{n}_{\sigma'}  \rangle\,.
\end{equation} 
The relativistic neutron operator, $\hat{n}$, can be written in terms
of its composition of $udd$ quarks, i.e.,
\begin{equation}
  \hat{n}_\sigma = \epsilon^{\alpha\beta\gamma}\,
  (\hat{u}^\top_{\alpha} {\cal C}\gamma_5 \hat{d}_{\beta}) \hat{d}_{\gamma \sigma}
  = \hat{\Phi}_{ud}^\gamma \hat{d}_{\gamma \sigma}\,,
\end{equation}
where we have introduced the ``good'' two-flavor diquark operator,
$\hat{\Phi}_{ud}^\gamma \equiv \epsilon^{\alpha\beta\gamma}
\hat{u}^\top_\alpha {\cal C}\gamma_5 \hat{d}_\beta$
[cf.\ Eqs.\,\eqref{eq:good} and \eqref{eq:2SCcond}].

It is then straightforward to rearrange the indices and factorize 
$\Upsilon_S \equiv  \langle\hat{\Upsilon}_S \rangle$ into
diquark condensates as
\begin{equation}
  \Upsilon_S = \langle\hat{\Phi}_{ud}^\alpha \hat{\Phi}_{ud}^\beta 
  \,\hat{d}^\top_\alpha {\cal C}\gamma_5 \hat{d}_\beta\rangle
  \approx \Phi_{ud}^\alpha\, \Phi_{ud}^\beta\, 
  \langle \hat{d}^\top_\alpha {\cal C}\gamma_5
  \hat{d}_\beta\rangle\,.
  \label{eq:ups}
\end{equation}
At high densities where the physical degrees of freedom are dominated
by quarks and the anti-triplet diquark condensate develops,
$\Upsilon_S$ should be largely given by the right-hand side in a sense
of a standard mean-field approximation;  we transform the
gauge-variant diquark field by introducing the fluctuation from its
mean value, and neglect the higher order fluctuation term.  Here, we
assumed unitary gauge fixing to make our discussion clear.
From the expression~\eqref{eq:ups} we see that a scalar $\langle dd\rangle$
condensate is induced in a scenario that smoothly connects 
superfluid neutron matter to quark matter.  The condensate
$\langle dd\rangle$ is symmetric in flavor and anti-symmetric in spin,
and hence symmetric in the color indices $\alpha$, $\beta$.  This
means that the permitted color structure belongs to the sextet
representation.  As previously argued in Sec.~\ref{sec:symmetry},
$\langle dd \rangle$ breaks $\mathrm{U(1)_B}$ and therefore exhibits
$S$-wave superfluidity. In essence, the neutron superfluid is
transformed continuously into the $d$-quark superfluid.

\subsection{Two-flavor $^3 P_2$ superfluid matter}

For $P$-wave superfluidity the quark-hadron continuity argument
proceeds in a similar way.  We start by writing down the pairing
operator of two neutrons in the $S=1$ and $L=1$ channel as
\begin{equation}
  \varphi_{\rm n}^\top \sigma^2 \sigma^i \nabla^j \varphi_{\rm n}\,,
\label{eq:order_p_nonrela}
\end{equation}
where the indices $i$ and $j$ run over spatial coordinates $x$, $y$,
and $z$.

Now, to address continuity from neutron matter to quark matter, we
need to generalize the pairing operator into a relativistic form.
This generalization may not be unique; some part of the spatial
derivative can emerge from the lower component of the
spinor\,\eqref{eq:non-rela}.  The only boundary condition is to
recover Eq.\,\eqref{eq:order_p_nonrela} in the non-relativistic limit,
and it is of course desirable to adopt expressions that are as simple
as possible.  One such candidate is
\begin{equation}
  \varphi_{\rm n}^\top \sigma^2 \sigma^i \nabla^j \varphi_{\rm n}
  \;\;\to\;\;
  \hat{\Upsilon}_P^{ij} \equiv \hat{n}^\top {\cal C} \gamma^i \nabla^j \hat{n}\,.
\end{equation}
With this operator, the index structures for the $^3 P_0$, $^3 P_1$,
and $^3 P_2$ channels can be further classified as follows:
\begin{align}
  ^3 P_0 : &\quad \hat{\Upsilon}_{P_0} = \hat{\Upsilon}_P^{ii} \,,\\
  ^3 P_1 : &\quad \hat{\Upsilon}_{P_1}^i = \epsilon^{ijk}\hat{\Upsilon}_P^{jk} \,,\\
  ^3 P_2 : &\quad \hat{\Upsilon}_{P_2}^{ij}
           = \hat{\Upsilon}_P^{ij} - \frac{1}{3} \delta^{ij}\,\hat{\Upsilon}_{P_0}\,.
\end{align}
The expression of the $^3 P_2$ operator above is in consonance with the
general form of the gap matrix for $J=2$ pairing~\cite{Bailin:1983bm}.
As we argued before, at sufficiently high baryon density the $^3 P_2$
state is favored.

Simple algebra as in the previous subsection then leads to the
rearrangement of the operators from neutrons to diquarks as follows:
\begin{align}
  & \Upsilon_P^{ij} = \langle \hat{\Upsilon}_P^{ij} \rangle \notag\\
  &\approx \Phi_{ud}^\alpha (\nabla^j \Phi_{ud}^\beta)
\langle \hat{d}^\top_\alpha {\cal C}\gamma^i \hat{d}_\beta\rangle
+ \Phi_{ud}^\alpha \Phi_{ud}^{\beta} \langle \hat{d}^\top_\alpha
{\cal C}\gamma^i \nabla^j \hat{d}_\beta \rangle\,.
\label{eq:Pwave}
\end{align}
Here the first term proportional to
$\boldsymbol{\nabla} \Phi_{ud}^\beta$ can be non-zero if the ground
state develops a crystalline color-superconducting phase in which the
Cooper pair carries a finite net momentum.  It is an interesting
problem how to optimize a possible interplay between the crystalline
profile and the spin-1 condensate $\langle d\gamma^i d\rangle$, but we
postpone this discussion and leave such a possibility for a future
study.

In this work we concentrate on the second term involving
$\langle d\gamma^i \nabla^j d\rangle$.  It is now evident that the
$^3P_2$ nature of neutron superfluidity is translated to that of $d$
quarks with their color configuration coupled to the scalar diquark
condensates in Eq.\,\eqref{eq:Pwave}.   As in the case of $^1 S_0$
superfluidity, this tensorial $\langle d\gamma^i \nabla^j d \rangle$
condensate would also retain the baryon superfluidity.  The symmetry
breaking patterns on both sides of neutron and quark matter become
exactly the same.  The remaining step is now to understand the
dynamics that favors $^3P_2$ over  $^1S_0$ pairing with increasing
baryon density.

\section{Dynamical properties favoring triplet $P$-wave pairing}
\label{sec:pwave}

Next we analyze dynamical mechanisms for $^3P_2$ pairing in the $dd$
channel.
This dynamical consideration is aimed to establish the
quark-hadron continuity and to match the quantum number of angular
momentum to the neutron superfluid, which also carries the ${}^3 P_2$
angular momentum quantum number.
We first discuss why $P$-wave pairing is preferred instead of
$S$-wave pairing.  Then the role of the spin-orbit interaction in
favoring the $J=2$ state among the $^3P_{J=0,1,2}$ channels will be
clarified.

\subsection{Short-range repulsive core favoring $L=1$}
\label{sec:NN}

Dense neutron matter is strongly affected by the short-distance
dynamics of the $NN$ interaction.  At low densities, the attractive
$^1S_0$ interaction dominates, while the $P$-wave ($L=1$) interaction
takes over at higher densities.  The short-range repulsion in the
$^1S_0$ channel acts to change the sign of the effective $nn$
interaction at the Fermi surface, from attractive to repulsive at
densities $\nB \gtrsim 0.5\,n_0$.  The $^1S_0$ pairing becomes
disfavored as compared to $P$-wave pairing.  The question is now
whether an analogous short-distance repulsive mechanism can be
identified in the interaction between two $d$-quarks.

At very high densities where a perturbative QCD treatment is feasible,
the quark scattering amplitude is well described by one-gluon exchange
with the following color structure:
\begin{align}
  \sum_{\mathcal{A} =1}^{8} T^{\mathcal{A}}_{\alpha\alpha'}
  T^{\mathcal{A}}_{\beta\beta'}
  = & - \frac13(\delta_{\alpha\alpha'} \delta_{\beta\beta'}
    - \delta_{\alpha\beta'} \delta_{\alpha'\beta}) \notag \\
  & + \frac16(\delta_{\alpha\alpha'} \delta_{\beta\beta'}
    + \delta_{\alpha\beta'} \delta_{\alpha'\beta})\,,
  \label{eq:oge}
\end{align}
where $T^{\mathcal{A}}$'s are the color SU(3) generators
($\mathcal{A} = 1,\ldots, 8$).  The first term corresponds to the
attractive $\bar{\boldsymbol{3}}_{A}$ channel, whereas the second term
corresponds to the repulsive $\boldsymbol{6}_{S}$ channel.  The color
structure of our $dd$ condensate is in fact in the symmetric color
sextet representation.  Therefore, in the perturbative region, the
short-range part of the interaction between $d$ quarks is repulsive
and naturally disfavors $S$-wave pairing.

In the confined phase, a short-distance repulsive interaction between
quarks can be thought of as emerging from quark-gluon exchange in a
non-relativistic  quark model picture (cf.\ the sketch in the middle
of Fig.\,\ref{fig:qhc}). Indeed it has been shown that the short-range
repulsive core in the $^1S_0$ channel of the nucleon-nucleon
interaction arises from the combined action of the Pauli principle and
the spin-spin force between
quarks~\cite{Oka:1980ax,Oka:1981ri,*Oka:1981rj}.  Using the resonating
group method, the scattering phase shifts between two nucleons in
$S$-wave were calculated and turned out to be negative (see Fig.\,2 of
Ref.\,\cite{Oka:1980ax}).  We show now that this mechanism correctly
accounts for the short-distance repulsive behavior of the interaction
between two $d$-quarks.

\begin{figure}
  \centering
  \includegraphics[width=0.4\columnwidth]{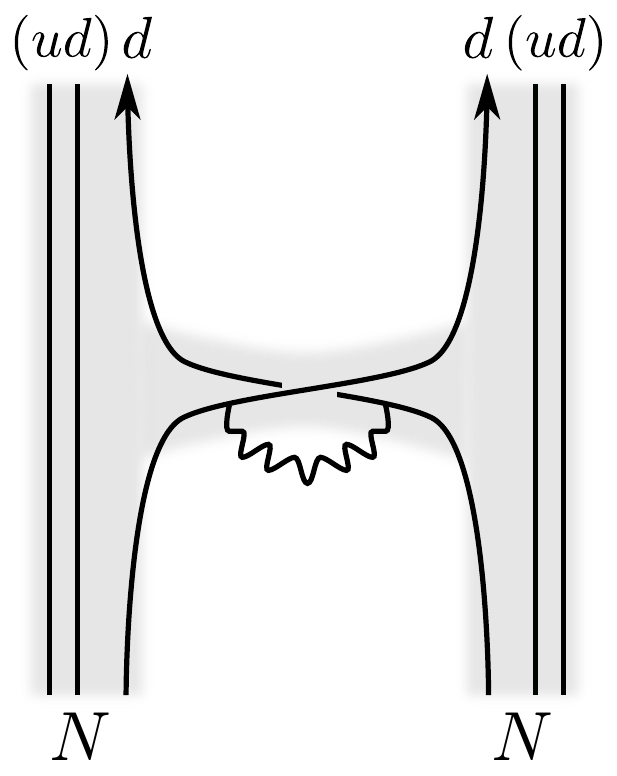}
  \caption{Short-range interaction between neutrons mediated by
    quark-gluon exchange.}
  \label{fig:NN}
\end{figure}

In the non-relativistic quark model analysis of the interaction
between two nucleons, one needs to consider only single quark exchange
with spin-spin correlation.  Two- or three-quark exchange processes
are redundant modulo exchange of the two nucleons.  For two
interacting neutrons, assuming $\langle ud \rangle$ pairing in 2SC
configurations (see Sec.~\ref{sec:two}), one can therefore focus on
the exchange interaction between the two $d$-quarks and construct the
$dd$ potential as illustrated in Fig.\,\ref{fig:NN}: two $d$ quarks
cross their lines in the presence of an exchanged gluon. Direct gluon
exchange without quark exchange is not allowed because of color
selection rules.  The one-gluon exchange (OGE) potential
reads~\cite{DeRujula:1975qlm}:
\begin{equation}
  V_{12}^{\rm OGE} = \left(\sum_{\mathcal{A}} T^{\mathcal{A}}_{1}
  T^{\mathcal{A}}_{2}\right) \frac{\alpha_s}4 \left[\frac1{r_{12}} -
    \frac{2\pi}{3m_q^2} (\boldsymbol{s}_1\!\cdot\!
    \boldsymbol{s}_2)\,\delta^3(\boldsymbol{r}_{12})\right]\,,
\end{equation}
omitting the tensor term in this expression.  Here
$\boldsymbol{r}_{12}$ denotes the distance between quarks 1 and
2.  Their spin operators are denoted by $\boldsymbol{s}_{1}$ and
$\boldsymbol{s}_{2}$.  The color structure in front of the potential
is exactly the same as the representation in Eq.~\eqref{eq:oge}.  In
close analogy with the $nn$ interaction, short-range repulsion appears
in the $dd$ potential.  Therefore pairing in $L=0$ is disfavored and
superfluidity appears predominantly in the $L=1$ state.

\subsection{Spin-orbit interaction favoring $J=2$}
\label{sec:LS}

The previous discussion has pointed to $dd$ quark pairing in $^3P_J$
states.  While the spin triplet necessarily follows in $L=1$ states
from the statistics of the wave function, the total angular momentum
$J$ is still left unspecified.

In neutron star matter, $^3P_2$ neutron superfluidity occurs because
of the strong spin-orbit interaction between neutrons.  The matrix
elements of
\begin{equation}
  \boldsymbol{L}\cdot\boldsymbol{S}
  = {1\over 2}\left[J(J+1) - L(L+1)-S(S+1)\right]
\end{equation}
are $-2, -1$ and $+1$ in $^3P_0$, $^3P_1$ and $^3P_2$ states,
respectively.  With an extra minus sign in the spin-orbit potential,
there is attraction in $^3P_2$ and repulsion in $^3P_{J=0,1}$
channels.  These features are also reflected in the empirical triplet
$P$-wave phase shifts.  The tensor force in $^3P_2$ is relatively
weak: ten times smaller than the one in the $^3P_0$ channel.  In the
absence of the spin-orbit force, superfluidity would in fact appear in
$^3P_0$.

The neutron-neutron spin-orbit interaction is generated by Lorentz
scalar and vector couplings of the neutrons. In chiral theories, such
couplings involve two- and three-pion exchange mechanisms.
Phenomenological boson exchange models
\cite{Nagels:1975fb, Machleidt:1987hj} associate these interactions
with scalar and vector boson fields, $\sigma(x)$ and $v^\mu(x)$.  The
vector field includes isoscalar and isovector terms (sometimes
identified with $\omega$ and $\rho$ mesons, but ultimately
representing multi-pion exchange mechanisms together with
short-distance dynamics). In the neutron-neutron interaction the
isoscalar and isovector terms have the same weight (the extra factor
in the isovector part is $\boldsymbol{\tau}_1\cdot\boldsymbol{\tau}_2
=1$).

We start from the following boson-nucleon vertex Lagrangians: 
\begin{equation}
  \begin{split}
{\cal L}_S =& - g_S\,\bar{\psi}(x)\psi(x)\,\sigma(x)\,, \\
{\cal L}_V =& - g_V\,\bar{\psi}(x)\gamma_\mu\psi(x)\,v^\mu(x) \\
&+ {g_T\over 2m_N}\bar{\psi}(x)\sigma_{\mu\nu}\psi(x)\,\partial^\nu   v^\mu(x)\,, 
\end{split}
\label{eq:Lagrangians}
\end{equation}
where $m_N$ is the nucleon mass.  Scalar and vector boson masses will
be denoted by $m_S$ and $m_V$.  Next, consider the momentum space
matrix elements of nucleon-nucleon $t$-channel Born terms generated by
these vertices and identify their spin-orbit pieces.  In the $NN$
center-of-mass frame, introduce initial and final state momenta,
$\boldsymbol{p}$ and $\boldsymbol{p'}$, and total spin
$\boldsymbol{S} = \boldsymbol{s_1} + \boldsymbol{s_2}$.  Furthermore,
\begin{equation}
\boldsymbol{P} = {1\over 2}(\boldsymbol{p} + \boldsymbol{p'})~,~~~\boldsymbol{q} = \boldsymbol{p'} - \boldsymbol{p}\,.
\end{equation}
The spin-orbit interaction matrix element deduced from interactions in
Eq.\,\eqref{eq:Lagrangians} to (leading) order
$\boldsymbol{p}^2/ m_N^2$ is:
\begin{align}
&\langle\boldsymbol{p'}|V_{LS}|\boldsymbol{p}\rangle =\notag\\
&-{i\over 2m_N^2}\left[{g_S^2\over \boldsymbol{q}^2 + m_S^2}+ {3g_V^2 + 4g_V g_T\over\boldsymbol{q}^2 + m_V^2} \right]\boldsymbol{S}\cdot(\boldsymbol{P}\times\boldsymbol{q})\,.
\label{eq:VLS}
\end{align}
We note that upon Fourier transformation, Eq.\,\eqref{eq:VLS} turns
into the $r$-space spin-orbit potential,
\begin{align}
V_{LS}(\boldsymbol{r}) &= {1\over 2m_N^2r}\,{d f(r)\over dr}\,\boldsymbol{L}\cdot\boldsymbol{S}\,,\notag\\
f(r) &= {g_S^2\over 4\pi}{e^{-m_S r}\over r}+ {g_V^2\over 4\pi}\left(3+{4g_T\over g_V}\right){e^{-m_V r}\over r}\,,
\end{align}
with $\boldsymbol{L} = \boldsymbol{r}\times\boldsymbol{P}$.  For
$\langle\boldsymbol{L\cdot S}\rangle = +1$ in the $^3P_2$ channel, the
spin-orbit potential is \textit{attractive} since
$d/dr(e^{-mr}/r) = -(1+mr)e^{-mr}/r^2 < 0$.

Let us make a quick estimate of the magnitude of the
$\boldsymbol{L}\cdot\boldsymbol{S}$ force at a distance $r\sim 1 \fm$
between two nucleons.  The isoscalar coupling parameters are, roughly,
$g_S^2/4\pi \sim 8$ together with $g_V\simeq g_S$ and $g_T \simeq
0$.  The isovector vector interaction has $g_V^2/4\pi\simeq 0.5$ and
$g_T/g_V\simeq 6$ (with contributions from isoscalar and isovector
vector interactions to be added in $I = 1$ states such as two
neutrons).  Using boson masses $m_S\sim 0.5$ GeV and $m_V\sim 0.8$
GeV, this gives in the neutron-neutron $^3P_2$ channel:
\begin{equation}
  V_{LS}^{nn}(r\sim 1\,\rm{fm}) \simeq -24 \MeV\,, 
\label{eq:LSnn}
\end{equation}
that is a characteristic order-of-magnitude documented by nuclear
phenomenology.  Recall that an average distance of about 1 fm between
nucleons corresponds to densities 5 -- 6 times the density of normal
nuclear matter, so it is already representative of the situation in
neutron star cores.

Now consider by analogy a corresponding scenario in terms of quarks,
i.e., we wish to examine $^3P_2$ superfluidity in the context of
hadron-quark continuity.  We seek possible mechanisms that generate an
$\boldsymbol{L} \cdot \boldsymbol{S}$ force at the quark level. 

The spin-orbit interaction between quarks can be
produced by one-gluon exchange:
\begin{align}
  \langle \bp' | V_{LS} | \bp \rangle
  =  \frac{-i}{2m_q^2}
  \left(\sum_{\mathcal{A}} T^{\mathcal{A}}_{1}
  T^{\mathcal{A}}_{2}\right)
  \frac{12\pi \alpha_s}{\boldsymbol{q}^2}
  \boldsymbol{S} \cdot (\boldsymbol{P} \times \boldsymbol{q}).
  \label{eq:ogels}
\end{align}
Fourier transforming this amplitude, one obtains the spin-orbit potential:
\begin{align}
  V_{LS}^{\rm OGE}(\boldsymbol{r})
  = -\frac{\alpha_s}{2 m_q^2 r^3} \boldsymbol{L} \cdot \boldsymbol{S},
\end{align}
where we have taken color $\boldsymbol{6}$ channel whose color
prefactor is $\sum_{\mathcal{A}}T_1^{\mathcal{A}}T_2^{\mathcal{A}} =
1/3$.  In a ${}^3 P_2$ state, the order of magnitude of a spin-orbit
attraction between two quarks exchanging a gluon is
\begin{align}
  V_{LS}^{\rm OGE}(r) = - 42.5\, \alpha_s \left(\frac{r}{\text{fm}}\right)^{-3} \left(\frac{m_q}{300\,\text{MeV}}\right)^{-2} \,\text{MeV}\,.
\end{align}
With $\alpha_s\sim 0.5$ we see that $V_{LS}^{\rm OGE}(r\sim
1\,\text{fm})$ is comparable to the aforementioned value of
Eq.~(\ref{eq:LSnn}).

Alternatively, consider NJL-type models that describe the
quasiparticle nature of quarks in the presence of spontaneously broken
chiral symmetry.  Such models have frequently been used in
extrapolations to high density matter.  We refer to a recent version
that includes scalar and vector couplings together with diquark
correlations~\cite{Baym:2019iky,Song:2019qoh}:
\begin{equation}
  {\cal L}_{\rm int} = G(\bar{q}q)^2 + H(\bar{q}\bar{q})(qq) - G_V(\bar{q}\gamma^\mu q)^2\,.
\label{eq:NJL}
\end{equation} 
This model is guided by the quark-hadron continuity hypothesis and
designed to meet the stiffness conditions on the EoS of dense matter
imposed by the existence of heavy (two-solar-mass) neutron stars and
gravitational wave signals from neutron star merger events.  It
features a strongly repulsive vector interaction with a coupling
strength $G_V$ comparable in magnitude to the scalar coupling $G$
which in turn produces a ``constituent'' quark mass of about 0.3 GeV
starting from almost massless $u$ and $d$ quarks.  Typical values of
coupling strengths are
\begin{equation}
  G \simeq 2\,\Lambda^{-2}~,\quad \Lambda\simeq 0.6\,{\rm GeV}~,\quad G_V = (0.6 - 1.3)\,G\,.
\end{equation}
In the following we shall use $G_V \simeq G$ for guidance.

The scalar and vector interactions in Eq.\,\eqref{eq:NJL} generate
spin-orbit forces between quarks.  In order to compare with the
previous discussion for neutrons, it is useful to associate the
characteristic NJL cutoff $\Lambda$ with a mass scale in a bosonized
version of Eq.\,\eqref{eq:NJL} involving scalar and vector boson
fields, $\sigma(x)$ and $v^\mu(x)$:
\begin{equation}
  \begin{split}
  {\cal L}_S& = - \tilde{g}_S\,\bar{q}(x)q(x)\,\sigma(x)\,,\\
  {\cal L}_V& = - \tilde{g}_V\,\bar{q}(x)\gamma_\mu q(x)\,v^\mu(x)\,.
  \end{split}
\label{eq:NJL2}
\end{equation}
For example, the scalar field satisfies 
\begin{equation}
  (\boldsymbol{\nabla}^2 - \Lambda^2)\sigma(x)
  = \tilde{g}_S\,\bar{q}(x)q(x)\,,
\end{equation}
so that $\sigma = -(\tilde{g}_S/\Lambda^2)\bar{q}q$ and
$G = \tilde{g}_S^2/\Lambda^2$ in the long-wavelength limit.  Writing
the scalar field as an expectation value plus a fluctuating part,
$\sigma(x) = \langle\sigma\rangle + s(x)$, it is the expectation value
$\langle\sigma\rangle =
-(\tilde{g}_S/\Lambda^2)\langle\bar{q}q\rangle$ that determines the
constituent quark mass through the NJL gap equation,
$m_q = -2G\langle\bar{q}q\rangle$, while the fluctuating part $s(x)$
propagates between quark sources and generates exchange interactions.

The spin-orbit interaction between quarks produced by the scalar and
vector couplings~\eqref{eq:NJL2} is:
\begin{equation}
\langle\boldsymbol{p'}|V_{LS}|\boldsymbol{p}\rangle =
-{i\over 2m_q^2}\left[{\tilde{g}_S^2 + 3\tilde{g}_V^2\over\boldsymbol{q}^2 + \Lambda^2} \right]\boldsymbol{S}\cdot(\boldsymbol{P}\times\boldsymbol{q})\,. 
\label{eq:VLS2}
\end{equation}
By comparison with Eq.\,\eqref{eq:VLS} it becomes evident that the
spin-orbit forces between two neutrons and between two $d$-quarks are
of the same order of magnitude: with inclusion of the isoscalar vector
coupling in the neutron case (i.e., omitting the isovector vector
interaction for which there is no obvious NJL counterpart) we have,
roughly,
\begin{equation}
  {\tilde{g}_S^2 + 3\tilde{g}_V^2\over m_q^2\,\Lambda^2}
  \sim {g_S^2 + 3g_V^2\over m_N^2\,m_V^2}\,.
\end{equation}
This correspondence can be further illustrated by converting
Eq.\,\eqref{eq:VLS2} into an equivalent spin-orbit potential in
$r$-space, now operating between constituent quarks:
\begin{equation}
  \begin{split}
V_{LS}^{qq}(\boldsymbol{r}) &= {1\over 2m_q^2\,r}{d f(r)\over dr}\,\boldsymbol{L}\cdot\boldsymbol{S}\,,\\
f(r) &= {\left(\tilde{g}_S^2 + 3\tilde{g}_V^2\right)e^{-\Lambda r}\over 4\pi\,r}\,.
\end{split}
\label{eq:VLSqq}
\end{equation}
For example, two $d$-quarks in a $^3P_2$ state and at a distance
$r \sim 0.8$ fm experience a spin-orbit attraction of
\begin{equation}
  V_{LS}^{dd}(r \sim 0.8\,\text{fm}) \simeq -16\,\text{MeV}\,,
  \label{eq:LSdd}
\end{equation}
to be compared with Eq.\,\eqref{eq:LSnn}.  The values in
the $^3P_0$ and $^3P_1$ channels are +32\,MeV and +16\,MeV,
respectively. Correspondingly larger magnitudes for the spin-orbit potential
result if one takes the stronger vector coupling, $G_V = 1.3 \,G$ instead of 
$G_V = G$.

We can conclude that spin-orbit interactions between nucleons have a
close correspondence to spin-orbit interactions between quark
quasiparticles emerging from NJL-type models with strong vector
couplings.  One can also see that spin-orbit interactions between
quarks arising from one-gluon exchange reach a comparable magnitude.
As a consequence, the $^3P_2$ neutron superfluidity scenario in
neutron star matter has an analogue in a similarly favored $^3P_2$
superfluid pairing of $d$-quarks at high baryon densities.

\section{Supporting arguments}
\label{sec:obs}

We are proposing a novel phase, 2SC+$\langle dd\rangle$, which
inevitably arises from the viewpont of continuity to superfluid
neutron matter.
The existence of such an additional component $\langle dd\rangle$ is
suggested by further independent arguments.  Here we discuss the
rearrangement of diquark interaction terms and the neutron star
cooling phenomenology.

\subsection{Coupling to the energy-momentum tensor}
\label{sec:emt}

In the context of previous mean-field calculations of
color-superconducting quark matter in an NJL-type model (see, e.g.,
Refs.~\cite{Buballa:2003qv, Alford:2007xm} for a review),
a four-fermion coupling in the $^3P_2$ channel has so far been
missing.  It would then be instructive to see how the interaction in
this channel could be enlarged through the coupling to the
energy-momentum tensor in an explicit manner.  Let us consider the
four-fermion coupling in the $^3 P_2$ diquark channel, i.e.,
\begin{align}
  \hat{\calI}_P &= (\bar{\psi} \gamma^{i} \nabla^{j} \calC \bar{\psi}^\top)
            (\psi^\top \calC \gamma_{i} \nabla_{j} \psi) \notag\\
  &= (\gamma^{i} {\cal C})_{\sigma\sigma'} ({\cal C} \gamma_{i})_{\tau'\tau} \, \bar{\psi}_{\sigma} 
    (\nabla^{j} \psi)_{\tau} (\nabla_{j} \bar{\psi})_{\sigma'} \psi_{\tau'}\,,
\label{eq:IP-diquark}
\end{align}
where $\sigma,\tau,\dots$ are spin indices.

Using the Fierz transformation matrix given explicitly in the
Appendix~\ref{sec:fierz}, the Fierz-rearranged four-fermion coupling
is found in the form
\begin{align}
  & \hat{\calI}_P
  = - \frac{3}{4} (\bar{\psi} \nabla^{j} \psi)^{2}
  - \frac{3}{4} (\bar{\psi} \gamma^{0} \nabla^{j} \psi)^{2}
  - \frac{1}{4} (\bar{\psi} \gamma^{i} \nabla^{j} \psi)^{2} \notag\\
  &+ \frac{1}{4} (\bar{\psi} \sigma^{i0} \nabla^{j} \psi)^{2}
  - \frac{1}{8} (\bar{\psi} \sigma^{ij} \nabla^{k} \psi)^{2}
  + \frac{3}{4} (\bar{\psi} \gamma^{0} \gamma^{5} \nabla^{j} \psi)^{2} \notag\\
  &+ \frac{1}{4} (\bar{\psi} \gamma^{i} \gamma^{5} \nabla^{j} \psi)^{2}
  - \frac{3}{4} (\bar{\psi} i \gamma^{5} \nabla^{j} \psi)^{2}\,,
\label{eq:fierz}
\end{align}
where we have introduced the compact notation
$(\bar{\psi} \, \Gamma \,{\nabla}^{j} \psi)^2$ for
$(\bar{\psi} \, \Gamma \,\overrightarrow{\nabla}^{j} \psi)
(\bar{\psi}\, \overleftarrow{\nabla}_{j}\, \Gamma\, \psi)$ in each of
the terms on the right-hand side.

Notably, this Fierz transformed $\hat{\calI}_P$ has a direct
correspondence to the energy-momentum tensor in the fermionic sector,
$T^{\mu\nu}=\bar{\psi}i\gamma^\mu\partial^\nu\psi$.  For matter in
equilibrium, $T^{\mu\nu}=\diag[\varepsilon,-p,-p,-p]$, with the energy
density $\varepsilon$ and the pressure $p$ of fermionic matter.  The
tree-level expectation value of $\hat{\calI}_P$ in
Eq.\,\eqref{eq:fierz} thus becomes
\begin{equation}
  \langle\hat{\calI}_P\rangle \approx \frac{3}{4}p^2 \,.
  \label{eq:vevint}
\end{equation}
It is evident from this algebraic exercise that the $^3 P_2$ diquark
interaction couples to the pressure which is a macroscopic quantity.
Even if the direct mixing between the quark-antiquark (hole) and the
diquark sectors may not be large, the superfluid energy gap can be
enhanced by the macroscopic expectation value of the energy-momentum
tensor as given in Eq.\,\eqref{eq:vevint}.  Here, we also make a
remark about a gauge-invariant description of the $^3P_2$ diquark
condensate.  To form a gauge-invariant quantity the color indices are
saturated, and as long as $\langle\hat{\calI}_P\rangle\neq 0$ as in
Eq.\,\eqref{eq:vevint} and the quark-hadron continuity is postulated,
the $^3P_2$ diquark condensate squared is always mixed with the
energy-momentum tensor squared through $\langle\hat{\calI}_P\rangle\neq
0$.

\subsection{Aspects of neutron star cooling phenomenology}
\label{sec:2scx}

The temperature of a neutron star and its time evolution (cooling),
which can be read off from the thermal radiation from the stellar
surface, provides information about processes occurring in the
interior.
A salient feature of the mechanisms behind neutron star
cooling is their sensitivity to  possible quark degrees of freedom
inside the stellar core.

In attempts to describe the actual neutron star cooling data, pure 2SC
quark matter turns out not to work~\cite{Grigorian:2004jq}.  This is
due to the onset of the direct quark Urca process, which strongly affects
the cooling curve of the star.  If we assume pure 2SC matter only,
some $d$-quarks are not paired and remain as a normal component.  Thus
these residual quarks in the normal phase emit neutrinos via the
direct Urca process and efficiently induce cooling of the star.
Once the stellar mass exceeds a critical value for which the direct
Urca process sets in, the star cools too fast. 
In the earlier study of Ref.~\cite{Grigorian:2004jq}, one could in
principle explain the existing data (at the time of 2005) with pure
2SC matter, but it needs unlikely assumptions.
Hence, the existing data is in favor of
forming the pairing among the residual quarks, so that the superfluid
gap exponentially suppresses the direct Urca process.  A way to
proceed phenomenologically is to introduce  ad hoc an additional
species X with a hypothetical density-dependent pairing gap,
$\Delta_{\rm X}$, so that ``2SC + X'' matter fits the empirically
observed cooling pattern~\cite{Grigorian:2004jq}, including young and
cold sources such as PSR J0205+64 as well as old and warm sources such
as PSR 1055-52.  This additional weak pairing channel needs to have a
small gap $\Delta_{\rm X}$ ranging between $10 \keV$ and  $1 \MeV$.

One can speculate that $^1S_0$ or $^3P_2$ superfluidity of $d$-quarks
with its gap proportional to $\langle dd \rangle$ might be a natural
candidate to substitute for the unknown X{}.  The typical magnitude of
the neutron $^3P_2$ gap is $\Delta_{nn} \sim 0.1$
MeV~\cite{Baldo:1998ca,Ding:2016oxp} (see also the recent
review~\cite{Sedrakian:2018ydt}).  As we have pointed out in
Sec.\,\ref{sec:LS}, the attractive components of spin-orbit forces
between two neutrons or two $d$-quarks are of similar magnitude, so
that one can expect a gap, $\Delta_{dd}$ of order 10 - 100 keV, also
for $d$-quark pairing.  This would be in accordance with the
postulated properties of X{}.  Further justification by calculating
$\Delta_{dd}$ microscopically is left for future studies.  As
mentioned in Sec.~\ref{sec:two}, there may be ungapped modes, which
might dominate the direct Urca cooling. These modes, however, could be
heavy and eliminated from the low energy excitations by chiral symmetry
breaking.  A dynamical study of chiral aspects of $\langle dd
\rangle$ would also be motivated in this context.

\section{Summary and Conclusions}
\label{sec:summary}

Quark-hadron continuity postulates a soft crossover from hadronic to
quark degrees of freedom in  cold and dense baryonic matter if the
symmetry-breaking patterns in the hadronic and quark domains are
identical.  Under these conditions there is no phase transition from
hadrons to quarks.  This scenario has been rigorously formulated for
the idealized case of matter composed of three massless $(u,d,s)$
quark flavors.  The special situation with $N_F = N_C = 3$  implies
color-flavor locked (CFL) configurations of diquark condensates.  The
CFL phase of quark matter has the same symmetry-breaking pattern as
the corresponding three-flavor hadronic phase with a baryonic
superfluid.  As part of this joint pattern, chiral symmetry is
spontaneously broken in both hadronic and quark phases.

Explicit chiral symmetry breaking by the non-zero quark masses in QCD
separates the heavier strange quark from the light $u$ and $d$
quarks.  The composition of cold matter in the real world is therefore
governed by $u$ and $d$ quarks with their approximate isospin
symmetry.  Matter exists in the form of nuclei, and in the form of
neutron stars at higher baryon densities.  Idealized
three-flavor matter is not the preferred ground state.
The strange matter hypothesis is not ruled out here, but given the empirical 
stiffness constraints for the neutron star equation-of-state, we
relegate its possibility to even higher density scales.
One can then raise the question whether matter with two-flavor
symmetry is still characterized by quark-hadron continuity, or whether
the symmetry breaking patterns in hadronic matter versus quark matter
are fundamentally different so that they are separated by a phase
transition.

The present work addresses this issue for the case of neutron matter
and comes to the conclusion that quark-hadron continuity can indeed be
realized in such a two-flavor system.  The key to this conclusion
comes from a detailed investigation of superfluidity in both hadronic
and quark phases.  Dense matter in the core of neutron stars serves as
a prototype example.

In neutron matter at low densities, the attractive $S$-wave
interaction between neutrons at the Fermi surface generates $^1S_0$
superfluidity.  At higher densities the $S$-wave interaction turns
repulsive and $^3P_2$ neutron superfluidity takes over, driven by the
attractive spin-orbit interaction in this channel.  The prime question
from the viewpoint of quark-hadron continuity is whether, at even
higher densities, $^3P_2$ superfluidity has an analogue in quark
matter such that the associated order parameter can be translated
continuously from one phase of matter to the other. In the preceding
sections of this paper we have explored symmetry aspects and dynamical
mechanisms related to this issue. The basic results are the following:
\begin{itemize}
\item{Formal rearrangements including all relevant symmetries permit a
    systematic translation from dibaryonic operators to diquark
    operators and their respective condensates, for both two- and
    three-flavor symmetric matter.}
\item{For the interesting case of neutron matter, it is shown that
    superfluidity involving neutron pairs, $\langle nn \rangle$,
    transforms into the superfluid pairing of $d$-quarks,
    $\langle dd\rangle$, together with the formation of
    $\langle ud\rangle$ diquark condensates.}
\item{The strong short-range repulsion in the interaction of two
    neutrons has an analogue in the repulsive short-distance force
    between two $d$-quarks.  This mechanism disfavors $S$-wave
    superfluidity of $d$-quarks at high density, in the same way as it
    disfavors $^1S_0$ neutron superfluidity at baryon densities larger
    than about half the density of normal nuclear matter.}
\item{The strong spin-orbit interaction between nucleons has an
    analogous counterpart in a corresponding
    $\boldsymbol{L}\cdot\boldsymbol{S}$ force in the quark sector,
    generated by one-gluon exchange or by vector couplings of quarks as they appear, for
    example, in extended Nambu--Jona-Lasinio models.  The spin-orbit
    forces between two neutrons as well as between two $d$-quarks are
    attractive in the triplet $P$-wave channel with total angular
    momentum $J=2$.  Therefore $^3P_2$ superfluidity in neutron matter
    finds its direct correspondence in $^3P_2$ superfluidity produced
    by $d$-quark pairing in quark matter.}
\end{itemize}

Altogether these findings suggest the presence of identical symmetry
breaking patterns, and hence quark-hadron continuity, in the
transition from neutron matter to two-flavor quark matter.  The new
element in this case is the continuity of $^3P_2$ superfluidity
between the hadronic and the quark phase.  The associated order
parameter involves a tensor combination of spin and momentum.  The
corresponding $^3P_2$ four-fermion coupling has not been considered in
previous quark matter studies.  It offers novel perspectives, such as
its close connection to the pressure component of the energy-momentum
tensor, a macroscopic quantity.  The $^3P_2$ superfluidity is also of
interest in the context of neutron star cooling.  Our continuity
scenario could be relevant to the QCD phase diagram with an isospin
imbalance.  An interesting possibility would be a continuity scenario
between the $^3 P_2$ superfluidity and crystalline
color-superconducting states. These and related topics are to be
explored in future studies.

\begin{acknowledgments}
We thank Toru~Kojo and Kei~Iida  for useful
discussions.
We are grateful to Mark~Alford for a careful reading of the manuscript
and giving insightful comments.
We thank Tetsuo~Hatsuda for his comments and bringing
Eq.~\eqref{eq:ogels} to our attention.
K.~F.\ was supported by Japan Society for the Promotion 
of Science (JSPS) KAKENHI Grant No.\ 18H01211 and No.\ 19K21874.
W.~W.\ gratefully acknowledges 
the hospitality extended to him at Department of Physics, The University of Tokyo. 
\end{acknowledgments}

\appendix
\section{Fierz transformation}
\label{sec:fierz}

The Fierz transformation matrix used in the derivations of relations
in Sec.\,\ref{sec:emt} is displayed here in its explicit form for
convenience of readers.  For further details, readers can consult
Appendix~A of Ref.\,\cite{Buballa:2003qv}.

\begin{widetext}
The relevant Fierz identity is given in a matrix form as
\begin{equation}
  \renewcommand{\arraystretch}{1.5}
  {\cal D} =
  \begin{pmatrix}
     -\frac{1}{4} &  \frac{1}{4} &  \frac{1}{4} &  \frac{1}{4} &  \frac{1}{4} &  \frac{1}{4} &  \frac{1}{4} &  \frac{1}{4} \\
     -\frac{1}{4} &  \frac{1}{4} & -\frac{1}{4} & -\frac{1}{4} &  \frac{1}{4} & -\frac{1}{4} & \frac{1}{4} & -\frac{1}{4} \\
     -\frac{3}{4} & -\frac{3}{4} & -\frac{1}{4} &  \frac{1}{4} & -\frac{1}{4} & \frac{3}{4} & \frac{1}{4} & -\frac{3}{4} \\
     -\frac{3}{4} & -\frac{3}{4} &  \frac{1}{4} & -\frac{1}{4} & -\frac{1}{4} & -\frac{3}{4} &  \frac{1}{4} & \frac{3}{4} \\
     -\frac{3}{4} &  \frac{3}{4} & -\frac{1}{4} & -\frac{1}{4} & -\frac{1}{4} &  \frac{3}{4} & -\frac{1}{4} &  \frac{3}{4} \\
      \frac{1}{4} &  \frac{1}{4} & -\frac{1}{4} &  \frac{1}{4} & -\frac{1}{4} & -\frac{1}{4} & \frac{1}{4} &  \frac{1}{4} \\
      \frac{3}{4} & -\frac{3}{4} & -\frac{1}{4} & -\frac{1}{4} &  \frac{1}{4} &  \frac{3}{4} &  \frac{1}{4} &  \frac{3}{4} \\
      \frac{1}{4} &  \frac{1}{4} &  \frac{1}{4} & -\frac{1}{4} & -\frac{1}{4} &  \frac{1}{4} &  \frac{1}{4} & -\frac{1}{4}
  \end{pmatrix}
  \calGam\,,
\end{equation}
where the diquark and the quark-antiquark (hole) interaction channels are
\begin{equation}
  \renewcommand{\arraystretch}{1.5}
  {\cal D} =
  \begin{pmatrix}
    {\cal C}_{\sigma\sigma'} {\cal C}_{\tau'\tau} \\
    (\gamma^{0}{\cal C})_{\sigma\sigma'} ({\cal C}\gamma^{0})_{\tau'\tau} \\
    (\gamma^{i}{\cal C})_{\sigma\sigma'} ({\cal C}\gamma_{i})_{\tau'\tau} \\
    (\sigma^{i0}{\cal C})_{\sigma\sigma'} ({\cal C}\sigma_{i0})_{\tau'\tau} \\
    \frac12 (\sigma^{ij}{\cal C})_{\sigma\sigma'} ({\cal C}\sigma_{ij})_{\tau'\tau} \\
    (\gamma^{0} \gamma^{5}{\cal C})_{\sigma\sigma'} ({\cal C}\gamma^{0} \gamma^{5})_{\tau'\tau} \\
    (\gamma^{i} \gamma^{5}{\cal C})_{\sigma\sigma'} ({\cal C}\gamma_{i} \gamma^{5})_{\tau'\tau} \\
    (i \gamma^{5}{\cal C})_{\sigma\sigma'} (i {\cal C} \gamma^{5})_{\tau'\tau}
  \end{pmatrix}\,,
  \qquad
  \calGam =
  \begin{pmatrix}
    (\boldsymbol{1})_{\sigma\tau} (\boldsymbol{1})_{\sigma'\tau'} \\
    (\gamma^{0})_{\sigma\tau} (\gamma^{0})_{\sigma'\tau'} \\
    (\gamma^{i})_{\sigma\tau} (\gamma_{i})_{\sigma'\tau'} \\
    (\sigma^{i0})_{\sigma\tau} (\sigma_{i0})_{\sigma'\tau'} \\
    \frac12 (\sigma^{ij})_{\sigma\tau} (\sigma_{ij})_{\sigma'\tau'} \\
    (\gamma^{0} \gamma^{5})_{\sigma\tau} (\gamma^{0} \gamma^{5})_{\sigma'\tau'} \\
    (\gamma^{i} \gamma^{5})_{\sigma\tau} (\gamma_{i} \gamma^{5})_{\sigma'\tau'} \\
    (i \gamma^{5})_{\sigma\tau} (i \gamma^{5})_{\sigma'\tau'}
  \end{pmatrix}\,.
\end{equation}
Taking the inverse of the above matrix, we can immediately derive an
identity to reexpress the diquark interaction in terms of the
quark-antiquark (hole) interaction.  In this way we can read the
matrix elements to derive a translation from
Eq.\,\eqref{eq:IP-diquark} to Eq.\,\eqref{eq:fierz}.
\end{widetext}

\bibliographystyle{apsrev4-1}
\bibliography{continuity}

\begin{thebibliography}{65}%
\makeatletter
\providecommand \@ifxundefined [1]{%
 \@ifx{#1\undefined}
}%
\providecommand \@ifnum [1]{%
 \ifnum #1\expandafter \@firstoftwo
 \else \expandafter \@secondoftwo
 \fi
}%
\providecommand \@ifx [1]{%
 \ifx #1\expandafter \@firstoftwo
 \else \expandafter \@secondoftwo
 \fi
}%
\providecommand \natexlab [1]{#1}%
\providecommand \enquote  [1]{``#1''}%
\providecommand \bibnamefont  [1]{#1}%
\providecommand \bibfnamefont [1]{#1}%
\providecommand \citenamefont [1]{#1}%
\providecommand \href@noop [0]{\@secondoftwo}%
\providecommand \href [0]{\begingroup \@sanitize@url \@href}%
\providecommand \@href[1]{\@@startlink{#1}\@@href}%
\providecommand \@@href[1]{\endgroup#1\@@endlink}%
\providecommand \@sanitize@url [0]{\catcode `\\12\catcode `\$12\catcode
  `\&12\catcode `\#12\catcode `\^12\catcode `\_12\catcode `\%12\relax}%
\providecommand \@@startlink[1]{}%
\providecommand \@@endlink[0]{}%
\providecommand \url  [0]{\begingroup\@sanitize@url \@url }%
\providecommand \@url [1]{\endgroup\@href {#1}{\urlprefix }}%
\providecommand \urlprefix  [0]{URL }%
\providecommand \Eprint [0]{\href }%
\providecommand \doibase [0]{http://dx.doi.org/}%
\providecommand \selectlanguage [0]{\@gobble}%
\providecommand \bibinfo  [0]{\@secondoftwo}%
\providecommand \bibfield  [0]{\@secondoftwo}%
\providecommand \translation [1]{[#1]}%
\providecommand \BibitemOpen [0]{}%
\providecommand \bibitemStop [0]{}%
\providecommand \bibitemNoStop [0]{.\EOS\space}%
\providecommand \EOS [0]{\spacefactor3000\relax}%
\providecommand \BibitemShut  [1]{\csname bibitem#1\endcsname}%
\let\auto@bib@innerbib\@empty
\bibitem [{\citenamefont {Sch{\" a}fer}\ and\ \citenamefont
  {Wilczek}(1999)}]{Schafer:1998ef}%
  \BibitemOpen
  \bibfield  {author} {\bibinfo {author} {\bibfnamefont {T.}~\bibnamefont
  {Sch{\" a}fer}}\ and\ \bibinfo {author} {\bibfnamefont {F.}~\bibnamefont
  {Wilczek}},\ }\href {\doibase 10.1103/PhysRevLett.82.3956} {\bibfield
  {journal} {\bibinfo  {journal} {Phys. Rev. Lett.}\ }\textbf {\bibinfo
  {volume} {82}},\ \bibinfo {pages} {3956} (\bibinfo {year} {1999})},\ \Eprint
  {http://arxiv.org/abs/hep-ph/9811473} {arXiv:hep-ph/9811473 [hep-ph]}
  \BibitemShut {NoStop}%
\bibitem [{\citenamefont {Alford}\ \emph {et~al.}(1999)\citenamefont {Alford},
  \citenamefont {Berges},\ and\ \citenamefont {Rajagopal}}]{Alford:1999pa}%
  \BibitemOpen
  \bibfield  {author} {\bibinfo {author} {\bibfnamefont {M.~G.}\ \bibnamefont
  {Alford}}, \bibinfo {author} {\bibfnamefont {J.}~\bibnamefont {Berges}}, \
  and\ \bibinfo {author} {\bibfnamefont {K.}~\bibnamefont {Rajagopal}},\ }\href
  {\doibase 10.1016/S0550-3213(99)00410-1} {\bibfield  {journal} {\bibinfo
  {journal} {Nucl. Phys.}\ }\textbf {\bibinfo {volume} {B558}},\ \bibinfo
  {pages} {219} (\bibinfo {year} {1999})},\ \Eprint
  {http://arxiv.org/abs/hep-ph/9903502} {arXiv:hep-ph/9903502 [hep-ph]}
  \BibitemShut {NoStop}%
\bibitem [{\citenamefont {Hatsuda}\ \emph {et~al.}(2006)\citenamefont
  {Hatsuda}, \citenamefont {Tachibana}, \citenamefont {Yamamoto},\ and\
  \citenamefont {Baym}}]{Hatsuda:2006ps}%
  \BibitemOpen
  \bibfield  {author} {\bibinfo {author} {\bibfnamefont {T.}~\bibnamefont
  {Hatsuda}}, \bibinfo {author} {\bibfnamefont {M.}~\bibnamefont {Tachibana}},
  \bibinfo {author} {\bibfnamefont {N.}~\bibnamefont {Yamamoto}}, \ and\
  \bibinfo {author} {\bibfnamefont {G.}~\bibnamefont {Baym}},\ }\href {\doibase
  10.1103/PhysRevLett.97.122001} {\bibfield  {journal} {\bibinfo  {journal}
  {Phys. Rev. Lett.}\ }\textbf {\bibinfo {volume} {97}},\ \bibinfo {pages}
  {122001} (\bibinfo {year} {2006})},\ \Eprint
  {http://arxiv.org/abs/hep-ph/0605018} {arXiv:hep-ph/0605018 [hep-ph]}
  \BibitemShut {NoStop}%
\bibitem [{\citenamefont {Abuki}\ \emph {et~al.}(2010)\citenamefont {Abuki},
  \citenamefont {Baym}, \citenamefont {Hatsuda},\ and\ \citenamefont
  {Yamamoto}}]{Abuki:2010jq}%
  \BibitemOpen
  \bibfield  {author} {\bibinfo {author} {\bibfnamefont {H.}~\bibnamefont
  {Abuki}}, \bibinfo {author} {\bibfnamefont {G.}~\bibnamefont {Baym}},
  \bibinfo {author} {\bibfnamefont {T.}~\bibnamefont {Hatsuda}}, \ and\
  \bibinfo {author} {\bibfnamefont {N.}~\bibnamefont {Yamamoto}},\ }\href
  {\doibase 10.1103/PhysRevD.81.125010} {\bibfield  {journal} {\bibinfo
  {journal} {Phys. Rev.}\ }\textbf {\bibinfo {volume} {D81}},\ \bibinfo {pages}
  {125010} (\bibinfo {year} {2010})},\ \Eprint {http://arxiv.org/abs/1003.0408}
  {arXiv:1003.0408 [hep-ph]} \BibitemShut {NoStop}%
\bibitem [{\citenamefont {Yamamoto}\ \emph {et~al.}(2007)\citenamefont
  {Yamamoto}, \citenamefont {Tachibana}, \citenamefont {Hatsuda},\ and\
  \citenamefont {Baym}}]{Yamamoto:2007ah}%
  \BibitemOpen
  \bibfield  {author} {\bibinfo {author} {\bibfnamefont {N.}~\bibnamefont
  {Yamamoto}}, \bibinfo {author} {\bibfnamefont {M.}~\bibnamefont {Tachibana}},
  \bibinfo {author} {\bibfnamefont {T.}~\bibnamefont {Hatsuda}}, \ and\
  \bibinfo {author} {\bibfnamefont {G.}~\bibnamefont {Baym}},\ }\href {\doibase
  10.1103/PhysRevD.76.074001} {\bibfield  {journal} {\bibinfo  {journal} {Phys.
  Rev.}\ }\textbf {\bibinfo {volume} {D76}},\ \bibinfo {pages} {074001}
  (\bibinfo {year} {2007})},\ \Eprint {http://arxiv.org/abs/0704.2654}
  {arXiv:0704.2654 [hep-ph]} \BibitemShut {NoStop}%
\bibitem [{\citenamefont {Hatsuda}\ \emph {et~al.}(2008)\citenamefont
  {Hatsuda}, \citenamefont {Tachibana},\ and\ \citenamefont
  {Yamamoto}}]{Hatsuda:2008is}%
  \BibitemOpen
  \bibfield  {author} {\bibinfo {author} {\bibfnamefont {T.}~\bibnamefont
  {Hatsuda}}, \bibinfo {author} {\bibfnamefont {M.}~\bibnamefont {Tachibana}},
  \ and\ \bibinfo {author} {\bibfnamefont {N.}~\bibnamefont {Yamamoto}},\
  }\href {\doibase 10.1103/PhysRevD.78.011501} {\bibfield  {journal} {\bibinfo
  {journal} {Phys. Rev.}\ }\textbf {\bibinfo {volume} {D78}},\ \bibinfo {pages}
  {011501} (\bibinfo {year} {2008})},\ \Eprint {http://arxiv.org/abs/0802.4143}
  {arXiv:0802.4143 [hep-ph]} \BibitemShut {NoStop}%
\bibitem [{\citenamefont {Alford}\ \emph {et~al.}(2019)\citenamefont {Alford},
  \citenamefont {Baym}, \citenamefont {Fukushima}, \citenamefont {Hatsuda},\
  and\ \citenamefont {Tachibana}}]{Alford:2018mqj}%
  \BibitemOpen
  \bibfield  {author} {\bibinfo {author} {\bibfnamefont {M.~G.}\ \bibnamefont
  {Alford}}, \bibinfo {author} {\bibfnamefont {G.}~\bibnamefont {Baym}},
  \bibinfo {author} {\bibfnamefont {K.}~\bibnamefont {Fukushima}}, \bibinfo
  {author} {\bibfnamefont {T.}~\bibnamefont {Hatsuda}}, \ and\ \bibinfo
  {author} {\bibfnamefont {M.}~\bibnamefont {Tachibana}},\ }\href {\doibase
  10.1103/PhysRevD.99.036004} {\bibfield  {journal} {\bibinfo  {journal} {Phys.
  Rev.}\ }\textbf {\bibinfo {volume} {D99}},\ \bibinfo {pages} {036004}
  (\bibinfo {year} {2019})},\ \Eprint {http://arxiv.org/abs/1803.05115}
  {arXiv:1803.05115 [hep-ph]} \BibitemShut {NoStop}%
\bibitem [{\citenamefont {Chatterjee}\ \emph {et~al.}(2019)\citenamefont
  {Chatterjee}, \citenamefont {Nitta},\ and\ \citenamefont
  {Yasui}}]{Chatterjee:2018nxe}%
  \BibitemOpen
  \bibfield  {author} {\bibinfo {author} {\bibfnamefont {C.}~\bibnamefont
  {Chatterjee}}, \bibinfo {author} {\bibfnamefont {M.}~\bibnamefont {Nitta}}, \
  and\ \bibinfo {author} {\bibfnamefont {S.}~\bibnamefont {Yasui}},\ }\href
  {\doibase 10.1103/PhysRevD.99.034001} {\bibfield  {journal} {\bibinfo
  {journal} {Phys. Rev.}\ }\textbf {\bibinfo {volume} {D99}},\ \bibinfo {pages}
  {034001} (\bibinfo {year} {2019})},\ \Eprint
  {http://arxiv.org/abs/1806.09291} {arXiv:1806.09291 [hep-ph]} \BibitemShut
  {NoStop}%
\bibitem [{\citenamefont {Cherman}\ \emph {et~al.}(2018)\citenamefont
  {Cherman}, \citenamefont {Sen},\ and\ \citenamefont
  {Yaffe}}]{Cherman:2018jir}%
  \BibitemOpen
  \bibfield  {author} {\bibinfo {author} {\bibfnamefont {A.}~\bibnamefont
  {Cherman}}, \bibinfo {author} {\bibfnamefont {S.}~\bibnamefont {Sen}}, \ and\
  \bibinfo {author} {\bibfnamefont {L.~G.}\ \bibnamefont {Yaffe}},\ }\href@noop
  {} {\  (\bibinfo {year} {2018})},\ \Eprint {http://arxiv.org/abs/1808.04827}
  {arXiv:1808.04827 [hep-th]} \BibitemShut {NoStop}%
\bibitem [{\citenamefont {Hirono}\ and\ \citenamefont
  {Tanizaki}(2019{\natexlab{a}})}]{Hirono:2018fjr}%
  \BibitemOpen
  \bibfield  {author} {\bibinfo {author} {\bibfnamefont {Y.}~\bibnamefont
  {Hirono}}\ and\ \bibinfo {author} {\bibfnamefont {Y.}~\bibnamefont
  {Tanizaki}},\ }\href {\doibase 10.1103/PhysRevLett.122.212001} {\bibfield
  {journal} {\bibinfo  {journal} {Phys. Rev. Lett.}\ }\textbf {\bibinfo
  {volume} {122}},\ \bibinfo {pages} {212001} (\bibinfo {year}
  {2019}{\natexlab{a}})},\ \Eprint {http://arxiv.org/abs/1811.10608}
  {arXiv:1811.10608 [hep-th]} \BibitemShut {NoStop}%
\bibitem [{\citenamefont {Hirono}\ and\ \citenamefont
  {Tanizaki}(2019{\natexlab{b}})}]{Hirono:2019oup}%
  \BibitemOpen
  \bibfield  {author} {\bibinfo {author} {\bibfnamefont {Y.}~\bibnamefont
  {Hirono}}\ and\ \bibinfo {author} {\bibfnamefont {Y.}~\bibnamefont
  {Tanizaki}},\ }\href@noop {} {\  (\bibinfo {year} {2019}{\natexlab{b}})},\
  \Eprint {http://arxiv.org/abs/1904.08570} {arXiv:1904.08570 [hep-th]}
  \BibitemShut {NoStop}%
\bibitem [{\citenamefont {McLerran}\ and\ \citenamefont
  {Pisarski}(2007)}]{McLerran:2007qj}%
  \BibitemOpen
  \bibfield  {author} {\bibinfo {author} {\bibfnamefont {L.}~\bibnamefont
  {McLerran}}\ and\ \bibinfo {author} {\bibfnamefont {R.~D.}\ \bibnamefont
  {Pisarski}},\ }\href {\doibase 10.1016/j.nuclphysa.2007.08.013} {\bibfield
  {journal} {\bibinfo  {journal} {Nucl. Phys.}\ }\textbf {\bibinfo {volume}
  {A796}},\ \bibinfo {pages} {83} (\bibinfo {year} {2007})},\ \Eprint
  {http://arxiv.org/abs/0706.2191} {arXiv:0706.2191 [hep-ph]} \BibitemShut
  {NoStop}%
\bibitem [{\citenamefont {Fukushima}\ and\ \citenamefont
  {Kojo}(2016)}]{Fukushima:2015bda}%
  \BibitemOpen
  \bibfield  {author} {\bibinfo {author} {\bibfnamefont {K.}~\bibnamefont
  {Fukushima}}\ and\ \bibinfo {author} {\bibfnamefont {T.}~\bibnamefont
  {Kojo}},\ }\href {\doibase 10.3847/0004-637X/817/2/180} {\bibfield  {journal}
  {\bibinfo  {journal} {Astrophys. J.}\ }\textbf {\bibinfo {volume} {817}},\
  \bibinfo {pages} {180} (\bibinfo {year} {2016})},\ \Eprint
  {http://arxiv.org/abs/1509.00356} {arXiv:1509.00356 [nucl-th]} \BibitemShut
  {NoStop}%
\bibitem [{\citenamefont {McLerran}\ and\ \citenamefont
  {Reddy}(2019)}]{McLerran:2018hbz}%
  \BibitemOpen
  \bibfield  {author} {\bibinfo {author} {\bibfnamefont {L.}~\bibnamefont
  {McLerran}}\ and\ \bibinfo {author} {\bibfnamefont {S.}~\bibnamefont
  {Reddy}},\ }\href {\doibase 10.1103/PhysRevLett.122.122701} {\bibfield
  {journal} {\bibinfo  {journal} {Phys. Rev. Lett.}\ }\textbf {\bibinfo
  {volume} {122}},\ \bibinfo {pages} {122701} (\bibinfo {year} {2019})},\
  \Eprint {http://arxiv.org/abs/1811.12503} {arXiv:1811.12503 [nucl-th]}
  \BibitemShut {NoStop}%
\bibitem [{\citenamefont {Jeong}\ \emph {et~al.}(2019)\citenamefont {Jeong},
  \citenamefont {McLerran},\ and\ \citenamefont {Sen}}]{Jeong:2019lhv}%
  \BibitemOpen
  \bibfield  {author} {\bibinfo {author} {\bibfnamefont {K.~S.}\ \bibnamefont
  {Jeong}}, \bibinfo {author} {\bibfnamefont {L.}~\bibnamefont {McLerran}}, \
  and\ \bibinfo {author} {\bibfnamefont {S.}~\bibnamefont {Sen}},\ }\href@noop
  {} {\  (\bibinfo {year} {2019})},\ \Eprint {http://arxiv.org/abs/1908.04799}
  {arXiv:1908.04799 [nucl-th]} \BibitemShut {NoStop}%
\bibitem [{\citenamefont {Masuda}\ \emph {et~al.}(2013)\citenamefont {Masuda},
  \citenamefont {Hatsuda},\ and\ \citenamefont {Takatsuka}}]{Masuda:2012kf}%
  \BibitemOpen
  \bibfield  {author} {\bibinfo {author} {\bibfnamefont {K.}~\bibnamefont
  {Masuda}}, \bibinfo {author} {\bibfnamefont {T.}~\bibnamefont {Hatsuda}}, \
  and\ \bibinfo {author} {\bibfnamefont {T.}~\bibnamefont {Takatsuka}},\ }\href
  {\doibase 10.1088/0004-637X/764/1/12} {\bibfield  {journal} {\bibinfo
  {journal} {Astrophys. J.}\ }\textbf {\bibinfo {volume} {764}},\ \bibinfo
  {pages} {12} (\bibinfo {year} {2013})},\ \Eprint
  {http://arxiv.org/abs/1205.3621} {arXiv:1205.3621 [nucl-th]} \BibitemShut
  {NoStop}%
\bibitem [{\citenamefont {Alvarez-Castillo}\ \emph {et~al.}(2014)\citenamefont
  {Alvarez-Castillo}, \citenamefont {Benic}, \citenamefont {Blaschke},\ and\
  \citenamefont {\L{}astowiecki}}]{Alvarez-Castillo:2013spa}%
  \BibitemOpen
  \bibfield  {author} {\bibinfo {author} {\bibfnamefont {D.~E.}\ \bibnamefont
  {Alvarez-Castillo}}, \bibinfo {author} {\bibfnamefont {S.}~\bibnamefont
  {Benic}}, \bibinfo {author} {\bibfnamefont {D.}~\bibnamefont {Blaschke}}, \
  and\ \bibinfo {author} {\bibfnamefont {R.}~\bibnamefont {\L{}astowiecki}},\
  }\href {\doibase 10.5506/APhysPolBSupp.7.203} {\bibfield  {journal} {\bibinfo
   {journal} {Acta Phys. Polon. Supp.}\ }\textbf {\bibinfo {volume} {7}},\
  \bibinfo {pages} {203} (\bibinfo {year} {2014})},\ \Eprint
  {http://arxiv.org/abs/1311.5112} {arXiv:1311.5112 [nucl-th]} \BibitemShut
  {NoStop}%
\bibitem [{\citenamefont {Baym}\ \emph {et~al.}(2018)\citenamefont {Baym},
  \citenamefont {Hatsuda}, \citenamefont {Kojo}, \citenamefont {Powell},
  \citenamefont {Song},\ and\ \citenamefont {Takatsuka}}]{Baym:2017whm}%
  \BibitemOpen
  \bibfield  {author} {\bibinfo {author} {\bibfnamefont {G.}~\bibnamefont
  {Baym}}, \bibinfo {author} {\bibfnamefont {T.}~\bibnamefont {Hatsuda}},
  \bibinfo {author} {\bibfnamefont {T.}~\bibnamefont {Kojo}}, \bibinfo {author}
  {\bibfnamefont {P.~D.}\ \bibnamefont {Powell}}, \bibinfo {author}
  {\bibfnamefont {Y.}~\bibnamefont {Song}}, \ and\ \bibinfo {author}
  {\bibfnamefont {T.}~\bibnamefont {Takatsuka}},\ }\href {\doibase
  10.1088/1361-6633/aaae14} {\bibfield  {journal} {\bibinfo  {journal} {Rept.
  Prog. Phys.}\ }\textbf {\bibinfo {volume} {81}},\ \bibinfo {pages} {056902}
  (\bibinfo {year} {2018})},\ \Eprint {http://arxiv.org/abs/1707.04966}
  {arXiv:1707.04966 [astro-ph.HE]} \BibitemShut {NoStop}%
\bibitem [{\citenamefont {Baym}\ \emph {et~al.}(2019)\citenamefont {Baym},
  \citenamefont {Furusawa}, \citenamefont {Hatsuda}, \citenamefont {Kojo},\
  and\ \citenamefont {Togashi}}]{Baym:2019iky}%
  \BibitemOpen
  \bibfield  {author} {\bibinfo {author} {\bibfnamefont {G.}~\bibnamefont
  {Baym}}, \bibinfo {author} {\bibfnamefont {S.}~\bibnamefont {Furusawa}},
  \bibinfo {author} {\bibfnamefont {T.}~\bibnamefont {Hatsuda}}, \bibinfo
  {author} {\bibfnamefont {T.}~\bibnamefont {Kojo}}, \ and\ \bibinfo {author}
  {\bibfnamefont {H.}~\bibnamefont {Togashi}},\ }\href@noop {} {\  (\bibinfo
  {year} {2019})},\ \Eprint {http://arxiv.org/abs/1903.08963} {arXiv:1903.08963
  [astro-ph.HE]} \BibitemShut {NoStop}%
\bibitem [{\citenamefont {Fujimoto}\ \emph {et~al.}(2018)\citenamefont
  {Fujimoto}, \citenamefont {Fukushima},\ and\ \citenamefont
  {Murase}}]{Fujimoto:2017cdo}%
  \BibitemOpen
  \bibfield  {author} {\bibinfo {author} {\bibfnamefont {Y.}~\bibnamefont
  {Fujimoto}}, \bibinfo {author} {\bibfnamefont {K.}~\bibnamefont {Fukushima}},
  \ and\ \bibinfo {author} {\bibfnamefont {K.}~\bibnamefont {Murase}},\ }\href
  {\doibase 10.1103/PhysRevD.98.023019} {\bibfield  {journal} {\bibinfo
  {journal} {Phys. Rev.}\ }\textbf {\bibinfo {volume} {D98}},\ \bibinfo {pages}
  {023019} (\bibinfo {year} {2018})},\ \Eprint
  {http://arxiv.org/abs/1711.06748} {arXiv:1711.06748 [nucl-th]} \BibitemShut
  {NoStop}%
\bibitem [{\citenamefont {Fujimoto}\ \emph {et~al.}(2019)\citenamefont
  {Fujimoto}, \citenamefont {Fukushima},\ and\ \citenamefont
  {Murase}}]{Fujimoto:2019hxv}%
  \BibitemOpen
  \bibfield  {author} {\bibinfo {author} {\bibfnamefont {Y.}~\bibnamefont
  {Fujimoto}}, \bibinfo {author} {\bibfnamefont {K.}~\bibnamefont {Fukushima}},
  \ and\ \bibinfo {author} {\bibfnamefont {K.}~\bibnamefont {Murase}},\
  }\href@noop {} {\  (\bibinfo {year} {2019})},\ \Eprint
  {http://arxiv.org/abs/1903.03400} {arXiv:1903.03400 [nucl-th]} \BibitemShut
  {NoStop}%
\bibitem [{\citenamefont {{\"Ozel}}\ \emph {et~al.}(2010)\citenamefont
  {{\"Ozel}}, \citenamefont {Baym},\ and\ \citenamefont {Guver}}]{Ozel:2010fw}%
  \BibitemOpen
  \bibfield  {author} {\bibinfo {author} {\bibfnamefont {F.}~\bibnamefont
  {{\"Ozel}}}, \bibinfo {author} {\bibfnamefont {G.}~\bibnamefont {Baym}}, \
  and\ \bibinfo {author} {\bibfnamefont {T.}~\bibnamefont {Guver}},\ }\href
  {\doibase 10.1103/PhysRevD.82.101301} {\bibfield  {journal} {\bibinfo
  {journal} {Phys. Rev.}\ }\textbf {\bibinfo {volume} {D82}},\ \bibinfo {pages}
  {101301} (\bibinfo {year} {2010})},\ \Eprint {http://arxiv.org/abs/1002.3153}
  {arXiv:1002.3153 [astro-ph.HE]} \BibitemShut {NoStop}%
\bibitem [{\citenamefont {Ozel}\ \emph {et~al.}(2016)\citenamefont {Ozel},
  \citenamefont {Psaltis}, \citenamefont {Arzoumanian}, \citenamefont
  {Morsink},\ and\ \citenamefont {Baubock}}]{Ozel:2015ykl}%
  \BibitemOpen
  \bibfield  {author} {\bibinfo {author} {\bibfnamefont {F.}~\bibnamefont
  {Ozel}}, \bibinfo {author} {\bibfnamefont {D.}~\bibnamefont {Psaltis}},
  \bibinfo {author} {\bibfnamefont {Z.}~\bibnamefont {Arzoumanian}}, \bibinfo
  {author} {\bibfnamefont {S.}~\bibnamefont {Morsink}}, \ and\ \bibinfo
  {author} {\bibfnamefont {M.}~\bibnamefont {Baubock}},\ }\href {\doibase
  10.3847/0004-637X/832/1/92} {\bibfield  {journal} {\bibinfo  {journal}
  {Astrophys. J.}\ }\textbf {\bibinfo {volume} {832}},\ \bibinfo {pages} {92}
  (\bibinfo {year} {2016})},\ \Eprint {http://arxiv.org/abs/1512.03067}
  {arXiv:1512.03067 [astro-ph.HE]} \BibitemShut {NoStop}%
\bibitem [{\citenamefont {Steiner}\ \emph {et~al.}(2010)\citenamefont
  {Steiner}, \citenamefont {Lattimer},\ and\ \citenamefont
  {Brown}}]{Steiner:2010fz}%
  \BibitemOpen
  \bibfield  {author} {\bibinfo {author} {\bibfnamefont {A.~W.}\ \bibnamefont
  {Steiner}}, \bibinfo {author} {\bibfnamefont {J.~M.}\ \bibnamefont
  {Lattimer}}, \ and\ \bibinfo {author} {\bibfnamefont {E.~F.}\ \bibnamefont
  {Brown}},\ }\href {\doibase 10.1088/0004-637X/722/1/33} {\bibfield  {journal}
  {\bibinfo  {journal} {Astrophys. J.}\ }\textbf {\bibinfo {volume} {722}},\
  \bibinfo {pages} {33} (\bibinfo {year} {2010})},\ \Eprint
  {http://arxiv.org/abs/1005.0811} {arXiv:1005.0811 [astro-ph.HE]} \BibitemShut
  {NoStop}%
\bibitem [{\citenamefont {Steiner}\ \emph {et~al.}(2013)\citenamefont
  {Steiner}, \citenamefont {Lattimer},\ and\ \citenamefont
  {Brown}}]{Steiner:2012xt}%
  \BibitemOpen
  \bibfield  {author} {\bibinfo {author} {\bibfnamefont {A.~W.}\ \bibnamefont
  {Steiner}}, \bibinfo {author} {\bibfnamefont {J.~M.}\ \bibnamefont
  {Lattimer}}, \ and\ \bibinfo {author} {\bibfnamefont {E.~F.}\ \bibnamefont
  {Brown}},\ }\href {\doibase 10.1088/2041-8205/765/1/L5} {\bibfield  {journal}
  {\bibinfo  {journal} {Astrophys. J.}\ }\textbf {\bibinfo {volume} {765}},\
  \bibinfo {pages} {L5} (\bibinfo {year} {2013})},\ \Eprint
  {http://arxiv.org/abs/1205.6871} {arXiv:1205.6871 [nucl-th]} \BibitemShut
  {NoStop}%
\bibitem [{\citenamefont {Abbott}\ \emph {et~al.}(2018)\citenamefont {Abbott}
  \emph {et~al.}}]{Abbott:2018exr}%
  \BibitemOpen
  \bibfield  {author} {\bibinfo {author} {\bibfnamefont {B.~P.}\ \bibnamefont
  {Abbott}} \emph {et~al.} (\bibinfo {collaboration} {LIGO Scientific,
  Virgo}),\ }\href {\doibase 10.1103/PhysRevLett.121.161101} {\bibfield
  {journal} {\bibinfo  {journal} {Phys. Rev. Lett.}\ }\textbf {\bibinfo
  {volume} {121}},\ \bibinfo {pages} {161101} (\bibinfo {year} {2018})},\
  \Eprint {http://arxiv.org/abs/1805.11581} {arXiv:1805.11581 [gr-qc]}
  \BibitemShut {NoStop}%
\bibitem [{\citenamefont {Gezerlis}\ \emph {et~al.}(2015)\citenamefont
  {Gezerlis}, \citenamefont {Pethick},\ and\ \citenamefont
  {Schwenk}}]{Gezerlis:2014efa}%
  \BibitemOpen
  \bibfield  {author} {\bibinfo {author} {\bibfnamefont {A.}~\bibnamefont
  {Gezerlis}}, \bibinfo {author} {\bibfnamefont {C.~J.}\ \bibnamefont
  {Pethick}}, \ and\ \bibinfo {author} {\bibfnamefont {A.}~\bibnamefont
  {Schwenk}},\ }\href@noop {} {\bibfield  {journal} {\bibinfo  {journal} {in:
  Novel Superfluids, Vol. 2, Int. Series of Monographs in Physics, K.-H.
  Bennemann and J. B. Ketterson (eds.), Oxford University Press, Oxford}\
  }\textbf {\bibinfo {volume} {157}},\ \bibinfo {pages} {580} (\bibinfo {year}
  {2015})},\ \Eprint {http://arxiv.org/abs/1406.6109} {arXiv:1406.6109
  [nucl-th]} \BibitemShut {NoStop}%
\bibitem [{\citenamefont {Sedrakian}\ and\ \citenamefont
  {Clark}(2018)}]{Sedrakian:2018ydt}%
  \BibitemOpen
  \bibfield  {author} {\bibinfo {author} {\bibfnamefont {A.}~\bibnamefont
  {Sedrakian}}\ and\ \bibinfo {author} {\bibfnamefont {J.~W.}\ \bibnamefont
  {Clark}},\ }\href@noop {} {\  (\bibinfo {year} {2018})},\ \Eprint
  {http://arxiv.org/abs/1802.00017} {arXiv:1802.00017 [nucl-th]} \BibitemShut
  {NoStop}%
\bibitem [{\citenamefont {Hoffberg}\ \emph {et~al.}(1970)\citenamefont
  {Hoffberg}, \citenamefont {Glassgold}, \citenamefont {Richardson},\ and\
  \citenamefont {Ruderman}}]{Hoffberg:1970vqj}%
  \BibitemOpen
  \bibfield  {author} {\bibinfo {author} {\bibfnamefont {M.}~\bibnamefont
  {Hoffberg}}, \bibinfo {author} {\bibfnamefont {A.~E.}\ \bibnamefont
  {Glassgold}}, \bibinfo {author} {\bibfnamefont {R.~W.}\ \bibnamefont
  {Richardson}}, \ and\ \bibinfo {author} {\bibfnamefont {M.}~\bibnamefont
  {Ruderman}},\ }\href {\doibase 10.1103/PhysRevLett.24.775} {\bibfield
  {journal} {\bibinfo  {journal} {Phys. Rev. Lett.}\ }\textbf {\bibinfo
  {volume} {24}},\ \bibinfo {pages} {775} (\bibinfo {year} {1970})}\BibitemShut
  {NoStop}%
\bibitem [{\citenamefont {Tamagaki}(1970)}]{Tamagaki:1970ptp}%
  \BibitemOpen
  \bibfield  {author} {\bibinfo {author} {\bibfnamefont {R.}~\bibnamefont
  {Tamagaki}},\ }\href {\doibase 10.1143/PTP.44.905} {\bibfield  {journal}
  {\bibinfo  {journal} {Prog. Theor. Phys.}\ }\textbf {\bibinfo {volume}
  {44}},\ \bibinfo {pages} {905} (\bibinfo {year} {1970})}\BibitemShut
  {NoStop}%
\bibitem [{\citenamefont {Takatsuka}\ and\ \citenamefont
  {Tamagaki}(1993)}]{Takatsuka:1992ga}%
  \BibitemOpen
  \bibfield  {author} {\bibinfo {author} {\bibfnamefont {T.}~\bibnamefont
  {Takatsuka}}\ and\ \bibinfo {author} {\bibfnamefont {R.}~\bibnamefont
  {Tamagaki}},\ }\href {\doibase 10.1143/PTPS.112.27} {\bibfield  {journal}
  {\bibinfo  {journal} {Prog. Theor. Phys. Suppl.}\ }\textbf {\bibinfo {volume}
  {112}},\ \bibinfo {pages} {27} (\bibinfo {year} {1993})}\BibitemShut
  {NoStop}%
\bibitem [{\citenamefont {Bedaque}\ \emph {et~al.}(2003)\citenamefont
  {Bedaque}, \citenamefont {Rupak},\ and\ \citenamefont
  {Savage}}]{Bedaque:2003wj}%
  \BibitemOpen
  \bibfield  {author} {\bibinfo {author} {\bibfnamefont {P.~F.}\ \bibnamefont
  {Bedaque}}, \bibinfo {author} {\bibfnamefont {G.}~\bibnamefont {Rupak}}, \
  and\ \bibinfo {author} {\bibfnamefont {M.~J.}\ \bibnamefont {Savage}},\
  }\href {\doibase 10.1103/PhysRevC.68.065802} {\bibfield  {journal} {\bibinfo
  {journal} {Phys. Rev.}\ }\textbf {\bibinfo {volume} {C68}},\ \bibinfo {pages}
  {065802} (\bibinfo {year} {2003})},\ \Eprint
  {http://arxiv.org/abs/nucl-th/0305032} {arXiv:nucl-th/0305032 [nucl-th]}
  \BibitemShut {NoStop}%
\bibitem [{\citenamefont {Masuda}\ and\ \citenamefont
  {Nitta}(2016)}]{Masuda:2015jka}%
  \BibitemOpen
  \bibfield  {author} {\bibinfo {author} {\bibfnamefont {K.}~\bibnamefont
  {Masuda}}\ and\ \bibinfo {author} {\bibfnamefont {M.}~\bibnamefont {Nitta}},\
  }\href {\doibase 10.1103/PhysRevC.93.035804} {\bibfield  {journal} {\bibinfo
  {journal} {Phys. Rev.}\ }\textbf {\bibinfo {volume} {C93}},\ \bibinfo {pages}
  {035804} (\bibinfo {year} {2016})},\ \Eprint
  {http://arxiv.org/abs/1512.01946} {arXiv:1512.01946 [nucl-th]} \BibitemShut
  {NoStop}%
\bibitem [{\citenamefont {Watanabe}\ and\ \citenamefont
  {Pethick}(2017)}]{Watanabe:2017nzj}%
  \BibitemOpen
  \bibfield  {author} {\bibinfo {author} {\bibfnamefont {G.}~\bibnamefont
  {Watanabe}}\ and\ \bibinfo {author} {\bibfnamefont {C.~J.}\ \bibnamefont
  {Pethick}},\ }\href {\doibase 10.1103/PhysRevLett.119.062701} {\bibfield
  {journal} {\bibinfo  {journal} {Phys. Rev. Lett.}\ }\textbf {\bibinfo
  {volume} {119}},\ \bibinfo {pages} {062701} (\bibinfo {year} {2017})},\
  \Eprint {http://arxiv.org/abs/1704.08859} {arXiv:1704.08859 [nucl-th]}
  \BibitemShut {NoStop}%
\bibitem [{\citenamefont {Drischler}\ \emph {et~al.}(2017)\citenamefont
  {Drischler}, \citenamefont {{Kr\"uger}}, \citenamefont {Hebeler},\ and\
  \citenamefont {Schwenk}}]{Drischler:2016cpy}%
  \BibitemOpen
  \bibfield  {author} {\bibinfo {author} {\bibfnamefont {C.}~\bibnamefont
  {Drischler}}, \bibinfo {author} {\bibfnamefont {T.}~\bibnamefont
  {{Kr\"uger}}}, \bibinfo {author} {\bibfnamefont {K.}~\bibnamefont {Hebeler}},
  \ and\ \bibinfo {author} {\bibfnamefont {A.}~\bibnamefont {Schwenk}},\ }\href
  {\doibase 10.1103/PhysRevC.95.024302} {\bibfield  {journal} {\bibinfo
  {journal} {Phys. Rev.}\ }\textbf {\bibinfo {volume} {C95}},\ \bibinfo {pages}
  {024302} (\bibinfo {year} {2017})},\ \Eprint
  {http://arxiv.org/abs/1610.05213} {arXiv:1610.05213 [nucl-th]} \BibitemShut
  {NoStop}%
\bibitem [{\citenamefont {Kitazawa}\ \emph {et~al.}(2002)\citenamefont
  {Kitazawa}, \citenamefont {Koide}, \citenamefont {Kunihiro},\ and\
  \citenamefont {Nemoto}}]{Kitazawa:2003qmg}%
  \BibitemOpen
  \bibfield  {author} {\bibinfo {author} {\bibfnamefont {M.}~\bibnamefont
  {Kitazawa}}, \bibinfo {author} {\bibfnamefont {T.}~\bibnamefont {Koide}},
  \bibinfo {author} {\bibfnamefont {T.}~\bibnamefont {Kunihiro}}, \ and\
  \bibinfo {author} {\bibfnamefont {Y.}~\bibnamefont {Nemoto}},\ }\href
  {\doibase 10.1143/PTP.110.185, 10.1143/PTP.108.929} {\bibfield  {journal}
  {\bibinfo  {journal} {Prog. Theor. Phys.}\ }\textbf {\bibinfo {volume}
  {108}},\ \bibinfo {pages} {929} (\bibinfo {year} {2002})},\ \bibinfo {note}
  {[Erratum: Prog. Theor. Phys.110,no.1,185(2003)]},\ \Eprint
  {http://arxiv.org/abs/hep-ph/0207255} {arXiv:hep-ph/0207255 [hep-ph]}
  \BibitemShut {NoStop}%
\bibitem [{\citenamefont {Zhang}\ \emph {et~al.}(2009)\citenamefont {Zhang},
  \citenamefont {Fukushima},\ and\ \citenamefont {Kunihiro}}]{Zhang:2008ima}%
  \BibitemOpen
  \bibfield  {author} {\bibinfo {author} {\bibfnamefont {Z.}~\bibnamefont
  {Zhang}}, \bibinfo {author} {\bibfnamefont {K.}~\bibnamefont {Fukushima}}, \
  and\ \bibinfo {author} {\bibfnamefont {T.}~\bibnamefont {Kunihiro}},\ }\href
  {\doibase 10.1103/PhysRevD.79.014004} {\bibfield  {journal} {\bibinfo
  {journal} {Phys. Rev.}\ }\textbf {\bibinfo {volume} {D79}},\ \bibinfo {pages}
  {014004} (\bibinfo {year} {2009})},\ \Eprint {http://arxiv.org/abs/0808.0927}
  {arXiv:0808.0927 [hep-ph]} \BibitemShut {NoStop}%
\bibitem [{\citenamefont {Burkert}\ \emph {et~al.}(2018)\citenamefont
  {Burkert}, \citenamefont {Elouadrhiri},\ and\ \citenamefont
  {Girod}}]{Burkert:2018bqq}%
  \BibitemOpen
  \bibfield  {author} {\bibinfo {author} {\bibfnamefont {V.~D.}\ \bibnamefont
  {Burkert}}, \bibinfo {author} {\bibfnamefont {L.}~\bibnamefont
  {Elouadrhiri}}, \ and\ \bibinfo {author} {\bibfnamefont {F.~X.}\ \bibnamefont
  {Girod}},\ }\href {\doibase 10.1038/s41586-018-0060-z} {\bibfield  {journal}
  {\bibinfo  {journal} {Nature}\ }\textbf {\bibinfo {volume} {557}},\ \bibinfo
  {pages} {396} (\bibinfo {year} {2018})}\BibitemShut {NoStop}%
\bibitem [{\citenamefont {Fonseca}\ \emph {et~al.}(2016)\citenamefont {Fonseca}
  \emph {et~al.}}]{Fonseca:2016tux}%
  \BibitemOpen
  \bibfield  {author} {\bibinfo {author} {\bibfnamefont {E.}~\bibnamefont
  {Fonseca}} \emph {et~al.},\ }\href {\doibase 10.3847/0004-637X/832/2/167}
  {\bibfield  {journal} {\bibinfo  {journal} {Astrophys. J.}\ }\textbf
  {\bibinfo {volume} {832}},\ \bibinfo {pages} {167} (\bibinfo {year}
  {2016})},\ \Eprint {http://arxiv.org/abs/1603.00545} {arXiv:1603.00545
  [astro-ph.HE]} \BibitemShut {NoStop}%
\bibitem [{\citenamefont {Antoniadis}\ \emph {et~al.}(2013)\citenamefont
  {Antoniadis} \emph {et~al.}}]{Antoniadis:2013pzd}%
  \BibitemOpen
  \bibfield  {author} {\bibinfo {author} {\bibfnamefont {J.}~\bibnamefont
  {Antoniadis}} \emph {et~al.},\ }\href {\doibase 10.1126/science.1233232}
  {\bibfield  {journal} {\bibinfo  {journal} {Science}\ }\textbf {\bibinfo
  {volume} {340}},\ \bibinfo {pages} {6131} (\bibinfo {year} {2013})},\ \Eprint
  {http://arxiv.org/abs/1304.6875} {arXiv:1304.6875 [astro-ph.HE]} \BibitemShut
  {NoStop}%
\bibitem [{\citenamefont {Cromartie}\ \emph {et~al.}(2019)\citenamefont
  {Cromartie} \emph {et~al.}}]{Cromartie:2019kug}%
  \BibitemOpen
  \bibfield  {author} {\bibinfo {author} {\bibfnamefont {H.~T.}\ \bibnamefont
  {Cromartie}} \emph {et~al.},\ }\href {\doibase 10.1038/s41550-019-0880-2} {\
  (\bibinfo {year} {2019}),\ 10.1038/s41550-019-0880-2},\ \Eprint
  {http://arxiv.org/abs/1904.06759} {arXiv:1904.06759 [astro-ph.HE]}
  \BibitemShut {NoStop}%
\bibitem [{\citenamefont {Drews}\ and\ \citenamefont
  {Weise}(2017)}]{Drews:2016wpi}%
  \BibitemOpen
  \bibfield  {author} {\bibinfo {author} {\bibfnamefont {M.}~\bibnamefont
  {Drews}}\ and\ \bibinfo {author} {\bibfnamefont {W.}~\bibnamefont {Weise}},\
  }\href {\doibase 10.1016/j.ppnp.2016.10.002} {\bibfield  {journal} {\bibinfo
  {journal} {Prog. Part. Nucl. Phys.}\ }\textbf {\bibinfo {volume} {93}},\
  \bibinfo {pages} {69} (\bibinfo {year} {2017})},\ \Eprint
  {http://arxiv.org/abs/1610.07568} {arXiv:1610.07568 [nucl-th]} \BibitemShut
  {NoStop}%
\bibitem [{\citenamefont {Song}\ \emph {et~al.}(2019)\citenamefont {Song},
  \citenamefont {Baym}, \citenamefont {Hatsuda},\ and\ \citenamefont
  {Kojo}}]{Song:2019qoh}%
  \BibitemOpen
  \bibfield  {author} {\bibinfo {author} {\bibfnamefont {Y.}~\bibnamefont
  {Song}}, \bibinfo {author} {\bibfnamefont {G.}~\bibnamefont {Baym}}, \bibinfo
  {author} {\bibfnamefont {T.}~\bibnamefont {Hatsuda}}, \ and\ \bibinfo
  {author} {\bibfnamefont {T.}~\bibnamefont {Kojo}},\ }\href@noop {} {\
  (\bibinfo {year} {2019})},\ \Eprint {http://arxiv.org/abs/1905.01005}
  {arXiv:1905.01005 [astro-ph.HE]} \BibitemShut {NoStop}%
\bibitem [{\citenamefont {Wilczek}(2004)}]{Wilczek:2004im}%
  \BibitemOpen
  \bibfield  {author} {\bibinfo {author} {\bibfnamefont {F.}~\bibnamefont
  {Wilczek}},\ }in\ \href {\doibase 10.1142/9789812775344_0007} {\emph
  {\bibinfo {booktitle} {{From fields to strings: Circumnavigating theoretical
  physics. Ian Kogan memorial collection (3 volume set)}}}}\ (\bibinfo {year}
  {2004})\ pp.\ \bibinfo {pages} {322--338},\ \Eprint
  {http://arxiv.org/abs/hep-ph/0409168} {arXiv:hep-ph/0409168 [hep-ph]}
  \BibitemShut {NoStop}%
\bibitem [{\citenamefont {Jaffe}(2005)}]{Jaffe:2004ph}%
  \BibitemOpen
  \bibfield  {author} {\bibinfo {author} {\bibfnamefont {R.~L.}\ \bibnamefont
  {Jaffe}},\ }\href {\doibase 10.1016/j.physrep.2004.11.005} {\bibfield
  {journal} {\bibinfo  {journal} {Phys. Rept.}\ }\textbf {\bibinfo {volume}
  {409}},\ \bibinfo {pages} {1} (\bibinfo {year} {2005})},\ \Eprint
  {http://arxiv.org/abs/hep-ph/0409065} {arXiv:hep-ph/0409065 [hep-ph]}
  \BibitemShut {NoStop}%
\bibitem [{Note1()}]{Note1}%
  \BibitemOpen
  \bibinfo {note} {Strictly speaking, Eq.~\protect \textup {\hbox
  {\mathsurround \z@ \protect \normalfont (\ignorespaces \ref {eq:good}\unskip
  \@@italiccorr )}} is only valid for the good diquark. Two quarks in $\protect
  \boldsymbol {3}$ representation of color and flavor group reduce to $\protect
  \boldsymbol {3} \times \protect \boldsymbol {3} = \protect \mathaccentV
  {bar}016{\protect \boldsymbol {3}} + \protect \boldsymbol {6}$. The indices
  on the right-hand side of Eq.~\protect \textup {\hbox {\mathsurround \z@
  \protect \normalfont (\ignorespaces \ref {eq:good}\unskip \@@italiccorr )}}
  are those of $\protect \boldsymbol {3}$, while on the left-hand side they
  refer to $\protect \mathaccentV {bar}016{\protect \boldsymbol {3}}$ after the
  reduction. See also Eqs.~\protect \textup {\hbox {\mathsurround \z@ \protect
  \normalfont (\ignorespaces \ref {eq:CFLcond}\unskip \@@italiccorr )}} and
  \protect \textup {\hbox {\mathsurround \z@ \protect \normalfont
  (\ignorespaces \ref {eq:2SCcond}\unskip \@@italiccorr )}}.}\BibitemShut
  {Stop}%
\bibitem [{\citenamefont {Casalbuoni}\ and\ \citenamefont
  {Gatto}(1999)}]{Casalbuoni:1999wu}%
  \BibitemOpen
  \bibfield  {author} {\bibinfo {author} {\bibfnamefont {R.}~\bibnamefont
  {Casalbuoni}}\ and\ \bibinfo {author} {\bibfnamefont {R.}~\bibnamefont
  {Gatto}},\ }\href {\doibase 10.1016/S0370-2693(99)01032-1} {\bibfield
  {journal} {\bibinfo  {journal} {Phys. Lett.}\ }\textbf {\bibinfo {volume}
  {B464}},\ \bibinfo {pages} {111} (\bibinfo {year} {1999})},\ \Eprint
  {http://arxiv.org/abs/hep-ph/9908227} {arXiv:hep-ph/9908227 [hep-ph]}
  \BibitemShut {NoStop}%
\bibitem [{\citenamefont {Son}\ and\ \citenamefont
  {Stephanov}(2000{\natexlab{a}})}]{Son:1999cm}%
  \BibitemOpen
  \bibfield  {author} {\bibinfo {author} {\bibfnamefont {D.~T.}\ \bibnamefont
  {Son}}\ and\ \bibinfo {author} {\bibfnamefont {M.~A.}\ \bibnamefont
  {Stephanov}},\ }\href {\doibase 10.1103/PhysRevD.61.074012} {\bibfield
  {journal} {\bibinfo  {journal} {Phys. Rev.}\ }\textbf {\bibinfo {volume}
  {D61}},\ \bibinfo {pages} {074012} (\bibinfo {year} {2000}{\natexlab{a}})},\
  \Eprint {http://arxiv.org/abs/hep-ph/9910491} {arXiv:hep-ph/9910491 [hep-ph]}
  \BibitemShut {NoStop}%
\bibitem [{\citenamefont {Son}\ and\ \citenamefont
  {Stephanov}(2000{\natexlab{b}})}]{Son:2000tu}%
  \BibitemOpen
  \bibfield  {author} {\bibinfo {author} {\bibfnamefont {D.~T.}\ \bibnamefont
  {Son}}\ and\ \bibinfo {author} {\bibfnamefont {M.~A.}\ \bibnamefont
  {Stephanov}},\ }\href {\doibase 10.1103/PhysRevD.62.059902} {\bibfield
  {journal} {\bibinfo  {journal} {Phys. Rev.}\ }\textbf {\bibinfo {volume}
  {D62}},\ \bibinfo {pages} {059902} (\bibinfo {year} {2000}{\natexlab{b}})},\
  \Eprint {http://arxiv.org/abs/hep-ph/0004095} {arXiv:hep-ph/0004095 [hep-ph]}
  \BibitemShut {NoStop}%
\bibitem [{\citenamefont {Berges}\ and\ \citenamefont
  {Rajagopal}(1999)}]{Berges:1998rc}%
  \BibitemOpen
  \bibfield  {author} {\bibinfo {author} {\bibfnamefont {J.}~\bibnamefont
  {Berges}}\ and\ \bibinfo {author} {\bibfnamefont {K.}~\bibnamefont
  {Rajagopal}},\ }\href {\doibase 10.1016/S0550-3213(98)00620-8} {\bibfield
  {journal} {\bibinfo  {journal} {Nucl. Phys.}\ }\textbf {\bibinfo {volume}
  {B538}},\ \bibinfo {pages} {215} (\bibinfo {year} {1999})},\ \Eprint
  {http://arxiv.org/abs/hep-ph/9804233} {arXiv:hep-ph/9804233 [hep-ph]}
  \BibitemShut {NoStop}%
\bibitem [{\citenamefont {Huang}\ \emph {et~al.}(2002)\citenamefont {Huang},
  \citenamefont {Zhuang},\ and\ \citenamefont {Chao}}]{Huang:2001yw}%
  \BibitemOpen
  \bibfield  {author} {\bibinfo {author} {\bibfnamefont {M.}~\bibnamefont
  {Huang}}, \bibinfo {author} {\bibfnamefont {P.-f.}\ \bibnamefont {Zhuang}}, \
  and\ \bibinfo {author} {\bibfnamefont {W.-q.}\ \bibnamefont {Chao}},\ }\href
  {\doibase 10.1103/PhysRevD.65.076012} {\bibfield  {journal} {\bibinfo
  {journal} {Phys. Rev.}\ }\textbf {\bibinfo {volume} {D65}},\ \bibinfo {pages}
  {076012} (\bibinfo {year} {2002})},\ \Eprint
  {http://arxiv.org/abs/hep-ph/0112124} {arXiv:hep-ph/0112124 [hep-ph]}
  \BibitemShut {NoStop}%
\bibitem [{\citenamefont {Alford}\ \emph {et~al.}(2003)\citenamefont {Alford},
  \citenamefont {Bowers}, \citenamefont {Cheyne},\ and\ \citenamefont
  {Cowan}}]{Alford:2002rz}%
  \BibitemOpen
  \bibfield  {author} {\bibinfo {author} {\bibfnamefont {M.~G.}\ \bibnamefont
  {Alford}}, \bibinfo {author} {\bibfnamefont {J.~A.}\ \bibnamefont {Bowers}},
  \bibinfo {author} {\bibfnamefont {J.~M.}\ \bibnamefont {Cheyne}}, \ and\
  \bibinfo {author} {\bibfnamefont {G.~A.}\ \bibnamefont {Cowan}},\ }\href
  {\doibase 10.1103/PhysRevD.67.054018} {\bibfield  {journal} {\bibinfo
  {journal} {Phys. Rev.}\ }\textbf {\bibinfo {volume} {D67}},\ \bibinfo {pages}
  {054018} (\bibinfo {year} {2003})},\ \Eprint
  {http://arxiv.org/abs/hep-ph/0210106} {arXiv:hep-ph/0210106 [hep-ph]}
  \BibitemShut {NoStop}%
\bibitem [{\citenamefont {Alford}\ \emph {et~al.}(2014)\citenamefont {Alford},
  \citenamefont {Nishimura},\ and\ \citenamefont {Sedrakian}}]{Alford:2014doa}%
  \BibitemOpen
  \bibfield  {author} {\bibinfo {author} {\bibfnamefont {M.~G.}\ \bibnamefont
  {Alford}}, \bibinfo {author} {\bibfnamefont {H.}~\bibnamefont {Nishimura}}, \
  and\ \bibinfo {author} {\bibfnamefont {A.}~\bibnamefont {Sedrakian}},\ }\href
  {\doibase 10.1103/PhysRevC.90.055205} {\bibfield  {journal} {\bibinfo
  {journal} {Phys. Rev.}\ }\textbf {\bibinfo {volume} {C90}},\ \bibinfo {pages}
  {055205} (\bibinfo {year} {2014})},\ \Eprint {http://arxiv.org/abs/1408.4999}
  {arXiv:1408.4999 [hep-ph]} \BibitemShut {NoStop}%
\bibitem [{\citenamefont {Bailin}\ and\ \citenamefont
  {Love}(1984)}]{Bailin:1983bm}%
  \BibitemOpen
  \bibfield  {author} {\bibinfo {author} {\bibfnamefont {D.}~\bibnamefont
  {Bailin}}\ and\ \bibinfo {author} {\bibfnamefont {A.}~\bibnamefont {Love}},\
  }\href {\doibase 10.1016/0370-1573(84)90145-5} {\bibfield  {journal}
  {\bibinfo  {journal} {Phys. Rept.}\ }\textbf {\bibinfo {volume} {107}},\
  \bibinfo {pages} {325} (\bibinfo {year} {1984})}\BibitemShut {NoStop}%
\bibitem [{\citenamefont {Oka}\ and\ \citenamefont
  {Yazaki}(1980)}]{Oka:1980ax}%
  \BibitemOpen
  \bibfield  {author} {\bibinfo {author} {\bibfnamefont {M.}~\bibnamefont
  {Oka}}\ and\ \bibinfo {author} {\bibfnamefont {K.}~\bibnamefont {Yazaki}},\
  }\href {\doibase 10.1016/0370-2693(80)90046-5} {\bibfield  {journal}
  {\bibinfo  {journal} {Phys. Lett.}\ }\textbf {\bibinfo {volume} {90B}},\
  \bibinfo {pages} {41} (\bibinfo {year} {1980})}\BibitemShut {NoStop}%
\bibitem [{\citenamefont {Oka}\ and\ \citenamefont
  {Yazaki}(1981{\natexlab{a}})}]{Oka:1981ri}%
  \BibitemOpen
  \bibfield  {author} {\bibinfo {author} {\bibfnamefont {M.}~\bibnamefont
  {Oka}}\ and\ \bibinfo {author} {\bibfnamefont {K.}~\bibnamefont {Yazaki}},\
  }\href {\doibase 10.1143/PTP.66.556} {\bibfield  {journal} {\bibinfo
  {journal} {Prog. Theor. Phys.}\ }\textbf {\bibinfo {volume} {66}},\ \bibinfo
  {pages} {556} (\bibinfo {year} {1981}{\natexlab{a}})}\BibitemShut {NoStop}%
\bibitem [{\citenamefont {Oka}\ and\ \citenamefont
  {Yazaki}(1981{\natexlab{b}})}]{Oka:1981rj}%
  \BibitemOpen
  \bibfield  {author} {\bibinfo {author} {\bibfnamefont {M.}~\bibnamefont
  {Oka}}\ and\ \bibinfo {author} {\bibfnamefont {K.}~\bibnamefont {Yazaki}},\
  }\href {\doibase 10.1143/PTP.66.572} {\bibfield  {journal} {\bibinfo
  {journal} {Prog. Theor. Phys.}\ }\textbf {\bibinfo {volume} {66}},\ \bibinfo
  {pages} {572} (\bibinfo {year} {1981}{\natexlab{b}})}\BibitemShut {NoStop}%
\bibitem [{\citenamefont {De~Rujula}\ \emph {et~al.}(1975)\citenamefont
  {De~Rujula}, \citenamefont {Georgi},\ and\ \citenamefont
  {Glashow}}]{DeRujula:1975qlm}%
  \BibitemOpen
  \bibfield  {author} {\bibinfo {author} {\bibfnamefont {A.}~\bibnamefont
  {De~Rujula}}, \bibinfo {author} {\bibfnamefont {H.}~\bibnamefont {Georgi}}, \
  and\ \bibinfo {author} {\bibfnamefont {S.~L.}\ \bibnamefont {Glashow}},\
  }\href {\doibase 10.1103/PhysRevD.12.147} {\bibfield  {journal} {\bibinfo
  {journal} {Phys. Rev.}\ }\textbf {\bibinfo {volume} {D12}},\ \bibinfo {pages}
  {147} (\bibinfo {year} {1975})}\BibitemShut {NoStop}%
\bibitem [{\citenamefont {Nagels}\ \emph {et~al.}(1975)\citenamefont {Nagels},
  \citenamefont {Rijken},\ and\ \citenamefont {de~Swart}}]{Nagels:1975fb}%
  \BibitemOpen
  \bibfield  {author} {\bibinfo {author} {\bibfnamefont {M.~M.}\ \bibnamefont
  {Nagels}}, \bibinfo {author} {\bibfnamefont {T.~A.}\ \bibnamefont {Rijken}},
  \ and\ \bibinfo {author} {\bibfnamefont {J.~J.}\ \bibnamefont {de~Swart}},\
  }\href {\doibase 10.1103/PhysRevD.12.744} {\bibfield  {journal} {\bibinfo
  {journal} {Phys. Rev.}\ }\textbf {\bibinfo {volume} {D12}},\ \bibinfo {pages}
  {744} (\bibinfo {year} {1975})}\BibitemShut {NoStop}%
\bibitem [{\citenamefont {Machleidt}\ \emph {et~al.}(1987)\citenamefont
  {Machleidt}, \citenamefont {Holinde},\ and\ \citenamefont
  {Elster}}]{Machleidt:1987hj}%
  \BibitemOpen
  \bibfield  {author} {\bibinfo {author} {\bibfnamefont {R.}~\bibnamefont
  {Machleidt}}, \bibinfo {author} {\bibfnamefont {K.}~\bibnamefont {Holinde}},
  \ and\ \bibinfo {author} {\bibfnamefont {C.}~\bibnamefont {Elster}},\ }\href
  {\doibase 10.1016/S0370-1573(87)80002-9} {\bibfield  {journal} {\bibinfo
  {journal} {Phys. Rept.}\ }\textbf {\bibinfo {volume} {149}},\ \bibinfo
  {pages} {1} (\bibinfo {year} {1987})}\BibitemShut {NoStop}%
\bibitem [{\citenamefont {Buballa}(2005)}]{Buballa:2003qv}%
  \BibitemOpen
  \bibfield  {author} {\bibinfo {author} {\bibfnamefont {M.}~\bibnamefont
  {Buballa}},\ }\href {\doibase 10.1016/j.physrep.2004.11.004} {\bibfield
  {journal} {\bibinfo  {journal} {Phys. Rept.}\ }\textbf {\bibinfo {volume}
  {407}},\ \bibinfo {pages} {205} (\bibinfo {year} {2005})},\ \Eprint
  {http://arxiv.org/abs/hep-ph/0402234} {arXiv:hep-ph/0402234 [hep-ph]}
  \BibitemShut {NoStop}%
\bibitem [{\citenamefont {Alford}\ \emph {et~al.}(2008)\citenamefont {Alford},
  \citenamefont {Schmitt}, \citenamefont {Rajagopal},\ and\ \citenamefont
  {Sch{\"a}fer}}]{Alford:2007xm}%
  \BibitemOpen
  \bibfield  {author} {\bibinfo {author} {\bibfnamefont {M.~G.}\ \bibnamefont
  {Alford}}, \bibinfo {author} {\bibfnamefont {A.}~\bibnamefont {Schmitt}},
  \bibinfo {author} {\bibfnamefont {K.}~\bibnamefont {Rajagopal}}, \ and\
  \bibinfo {author} {\bibfnamefont {T.}~\bibnamefont {Sch{\"a}fer}},\ }\href
  {\doibase 10.1103/RevModPhys.80.1455} {\bibfield  {journal} {\bibinfo
  {journal} {Rev. Mod. Phys.}\ }\textbf {\bibinfo {volume} {80}},\ \bibinfo
  {pages} {1455} (\bibinfo {year} {2008})},\ \Eprint
  {http://arxiv.org/abs/0709.4635} {arXiv:0709.4635 [hep-ph]} \BibitemShut
  {NoStop}%
\bibitem [{\citenamefont {Grigorian}\ \emph {et~al.}(2005)\citenamefont
  {Grigorian}, \citenamefont {Blaschke},\ and\ \citenamefont
  {Voskresensky}}]{Grigorian:2004jq}%
  \BibitemOpen
  \bibfield  {author} {\bibinfo {author} {\bibfnamefont {H.}~\bibnamefont
  {Grigorian}}, \bibinfo {author} {\bibfnamefont {D.}~\bibnamefont {Blaschke}},
  \ and\ \bibinfo {author} {\bibfnamefont {D.}~\bibnamefont {Voskresensky}},\
  }\href {\doibase 10.1103/PhysRevC.71.045801} {\bibfield  {journal} {\bibinfo
  {journal} {Phys. Rev.}\ }\textbf {\bibinfo {volume} {C71}},\ \bibinfo {pages}
  {045801} (\bibinfo {year} {2005})},\ \Eprint
  {http://arxiv.org/abs/astro-ph/0411619} {arXiv:astro-ph/0411619 [astro-ph]}
  \BibitemShut {NoStop}%
\bibitem [{\citenamefont {Baldo}\ \emph {et~al.}(1998)\citenamefont {Baldo},
  \citenamefont {Elgaroey}, \citenamefont {Engvik}, \citenamefont
  {Hjorth-Jensen},\ and\ \citenamefont {Schulze}}]{Baldo:1998ca}%
  \BibitemOpen
  \bibfield  {author} {\bibinfo {author} {\bibfnamefont {M.}~\bibnamefont
  {Baldo}}, \bibinfo {author} {\bibfnamefont {O.}~\bibnamefont {Elgaroey}},
  \bibinfo {author} {\bibfnamefont {L.}~\bibnamefont {Engvik}}, \bibinfo
  {author} {\bibfnamefont {M.}~\bibnamefont {Hjorth-Jensen}}, \ and\ \bibinfo
  {author} {\bibfnamefont {H.~J.}\ \bibnamefont {Schulze}},\ }\href {\doibase
  10.1103/PhysRevC.58.1921} {\bibfield  {journal} {\bibinfo  {journal} {Phys.
  Rev.}\ }\textbf {\bibinfo {volume} {C58}},\ \bibinfo {pages} {1921} (\bibinfo
  {year} {1998})},\ \Eprint {http://arxiv.org/abs/nucl-th/9806097}
  {arXiv:nucl-th/9806097 [nucl-th]} \BibitemShut {NoStop}%
\bibitem [{\citenamefont {Ding}\ \emph {et~al.}(2016)\citenamefont {Ding},
  \citenamefont {Rios}, \citenamefont {Dussan}, \citenamefont {Dickhoff},
  \citenamefont {Witte}, \citenamefont {Polls},\ and\ \citenamefont
  {Carbone}}]{Ding:2016oxp}%
  \BibitemOpen
  \bibfield  {author} {\bibinfo {author} {\bibfnamefont {D.}~\bibnamefont
  {Ding}}, \bibinfo {author} {\bibfnamefont {A.}~\bibnamefont {Rios}}, \bibinfo
  {author} {\bibfnamefont {H.}~\bibnamefont {Dussan}}, \bibinfo {author}
  {\bibfnamefont {W.~H.}\ \bibnamefont {Dickhoff}}, \bibinfo {author}
  {\bibfnamefont {S.~J.}\ \bibnamefont {Witte}}, \bibinfo {author}
  {\bibfnamefont {A.}~\bibnamefont {Polls}}, \ and\ \bibinfo {author}
  {\bibfnamefont {A.}~\bibnamefont {Carbone}},\ }\href {\doibase
  10.1103/PhysRevC.94.029901, 10.1103/PhysRevC.94.025802} {\bibfield  {journal}
  {\bibinfo  {journal} {Phys. Rev.}\ }\textbf {\bibinfo {volume} {C94}},\
  \bibinfo {pages} {025802} (\bibinfo {year} {2016})},\ \bibinfo {note}
  {[Addendum: Phys. Rev.C94,no.2,029901(2016)]},\ \Eprint
  {http://arxiv.org/abs/1601.01600} {arXiv:1601.01600 [nucl-th]} \BibitemShut
  {NoStop}%
\end{thebibliography}%

\end{document}